\documentclass[aip, pof, preprint]{revtex4-1}

\usepackage{graphicx} 
\usepackage{dcolumn}
\usepackage{bm}
\usepackage{amssymb}
\usepackage{amsmath}
\usepackage{wasysym}
\usepackage{color}
\usepackage{hyperref}
\usepackage{dcolumn}
\usepackage{float}
\usepackage{flushend}
\usepackage{multirow}
\usepackage{cancel}
\hypersetup{
colorlinks,
citecolor=blue,
filecolor=blue,
linkcolor=blue,
urlcolor=blue}

\newcommand{\be}{\begin{equation}}
\newcommand{\ee}{\end{equation}} 
\newcommand{\bea}{\begin{eqnarray}}
\newcommand{\eea}{\end{eqnarray}}

\begin{document}


\title{{ Statistical features of rapidly rotating decaying turbulence: enstrophy and energy spectra, and coherent structures.}}
\author{Manohar K. Sharma}
\email{kmanohar@iitk.ac.in}
\affiliation{
 Department of Physics,
  Indian Institute of Technology Kanpur,
  Uttar Pradesh 208016, India
}
\author{Abhishek Kumar}
\email{ac7600@coventry.ac.uk}
\affiliation{
 Applied Mathematics Research Centre,
  Coventry University,
  Coventry CV15FB, The United Kingdom 
}
\author{Mahendra K. Verma}
\email{mkv@iitk.ac.in}
\affiliation{
 Department of Physics,
  Indian Institute of Technology Kanpur,
  Uttar Pradesh 208016, India
}
\author{Sagar Chakraborty}
\email{sagarc@iitk.ac.in}
\affiliation{
  Department of Physics,
  Indian Institute of Technology Kanpur,
  Uttar Pradesh 208016, India
}
%
%
\begin{abstract}
{ In this paper we investigate the properties of rapidly rotating decaying turbulence using numerical simulations and phenomenological modelling.  We find that as the turbulent flow evolves in time, the Rossby number decreases to $\sim 10^{-3}$,  and the flow becomes quasi-two-dimensional with strong coherent columnar structures arising due to the inverse cascade of  energy. We establish that a major fraction of  energy is confined in  Fourier modes $(\pm1,0,0)$ and $(0,\pm1,0)$ that correspond to the largest columnar structure in the flow. For wavenumbers ($k$) greater than the enstrophy dissipation wavenumber ($k_d$), our phenomenological arguments and numerical study show that the enstrophy flux and  spectrum  of a horizontal  cross-section perpendicular to the axis of rotation are given by $\epsilon_\omega\exp(-C(k/k_d)^2)$ and $C\epsilon_\omega^{2/3}k^{-1}\exp(-C(k/k_d)^2)$ respectively; for this 2D flow,  $\epsilon_\omega$ is the enstrophy dissipation rate, and $C$ is a constant. Using these results, we  propose a new form for the energy spectrum of  rapidly rotating decaying turbulence: $E(k)=C\epsilon_\omega^{2/3}k^{-3}\exp(-C(k/k_d)^2)$. This model of the energy spectrum is based on  wavenumber-dependent enstrophy flux, and it deviates significantly from  power law energy spectrum reported earlier. }

\end{abstract}
\maketitle

\section{Introduction}
\label{sec:intro}
One of the least understood non-equilibrium statistical mechanical system is a fully developed turbulent flow\cite{Goldenfeld:JSP2017,Ruelle:JSP2014,Chakraborty:PRE2009}. Kolmogorov\cite{Kolmogorov:DANS1941Structure,Kolmogorov:DANS1941Dissipation} proposed a theory for homogeneous and isotropic three-dimensional (3D) hydrodynamic turbulence, according  to which the  inertial-range energy spectrum $E(k) \sim \epsilon^{2/3} k^{-5/3}$.  Here $\epsilon$ is the energy dissipation rate that equals the energy flux, and $k$ is the wavenumber.  This theory successfully explains many experimental and numerical findings\cite{Davidson:book:Turbulence,Lesieur:book:Turbulence,MoninYaglom:book:v1,MoninYaglom:book:v2,Frisch:book,Leslie:book,Batchelor:book:Turbulence}. Kraichnan\cite{Kraichnan:PF1967_2D} however showed that the two-dimensional (2D) hydrodynamic turbulence has further complexities--here the small wavenumber Fourier modes exhibit inverse cascade of energy with  $E(k) \sim  k^{-5/3}$, while the large wavenumber Fourier modes exhibit forward enstrophy cascade with $E(k) \sim  k^{-3}$.   

The fluid flows in nature and in laboratory are generally quite complicated. For example,  they may involve  external magnetic field,  rotation, or buoyancy   \cite{Davidson:book:Turbulence,Lesieur:book:Turbulence,MoninYaglom:book:v1,MoninYaglom:book:v2,Frisch:book,Leslie:book,Batchelor:book:Turbulence}.  The aforementioned hydrodynamic turbulence phenomenologies play major role in modelling these flows.   Researchers have shown that the external magnetic field\cite{Verma:PR2004,Kraichnan:PF1965MHD,Iroshnikov:SA1964}
and  rotation\cite{Davidson:book:TurbulenceRotating,Lesieur:book:Turbulence,Sagaut:book} typically affect the energy spectrum in the inertial range.   In this paper we address the turbulence phenomenology of {rapidly rotating} decaying turbulence.  

Rotating flows are ubiquitous in nature, e.g., in ocean, atmosphere, celestial bodies, as well as in engineering applications.  In the rotating frame of reference, the flow is affected by the Coriolis and the centrifugal forces.  While the centrifugal force may be absorbed into the pressure gradient term of the Navier--Stokes equation, the Coriolis force, which is perpendicular to the direction of rotation, tends to make the flow quasi two-dimensional (2D). {The} Taylor--Proudman theorem\cite{Davidson:book:Turbulence}  predicts  formation of {the} Taylor columns and emergence of quasi  2D behaviour.  Note however that  the Taylor--Proudman theorem is applicable in  the linear limit of rapidly rotating steady flow.  The quasi-2D behaviour of the turbulent rotating fluid is however much more subtle and is full of defining signatures that are still not fully understood\cite{Chakraborty:EPJB2010,Chakraborty:EPL2007,Canuto:PF1997_rotation,Hossain:PF1994}. As mentioned earlier, 2D turbulence has its own caveats.   The velocity component along the rotation axis, although relatively weaker than the perpendicular component, plays a significant role in rotating turbulence. Additionally, one can not ignore the nonlinear advection term even when the fluid is rotating extremely fast. Thus such flows are quite delicate to model, and they are being studied vigorously. 

{Several models have been proposed to study the kinetic energy spectrum of  rotating turbulence.} Zeman\cite{Zeman:PF1994} proposed a dual spectrum: the large-$k$ modes exhibit Kolmogorov's  spectrum ($k^{-5/3}$), while small-$k$ modes  show $k^{-11/5}$ spectrum. {Zhou\cite{Zhou:PF1995} proposed that $E(k) \sim k^{-2}$ for the entire inertial range when the rotation rate is very high}, whereas Smith and Waleffe\cite{Smith:PF1999} argued that $E(k_\perp) \sim k_\perp^{-3}$, where $k_\perp, k_\parallel$ are respectively the components of  wavevector ${\bf k}$ perpendicular and parallel to the rotation axis. Chakraborty
\cite{Chakraborty:PRE2007} argued that $E(k)\sim k^{m}$ where $m\in[-2,-3]$;  the spectral range  can be further confined to $m\in[-2,-7/3]$ by kinetic helicity. Additionally, in the limit of very strong rotation, Kraichnan\cite{Kraichnan:JFM1959} proposed that $E(k)\sim\exp(-2\nu k^2 t)$, where $\nu$ is the kinematic viscosity and $t$ is the time elapsed.  { This model assumes absence of nonlinearity.  In this paper we show that the nonlinearity, though weak, is present, and it produces nonzero energy and enstrophy fluxes.}



{Many researchers~\cite{Godeferd:AMR2015,Castello:PRE2011,Bokhoven:PF2009,Davidson:JFM2006a,Hopfinger:JFM1982,Baroud:PRL2002}, have attempted to verify the aforementioned models of rotating turbulence using experiments.} Morize {\em et al.}\cite{Morize:PF2005} studied decaying  turbulence in a rotating tank  and showed that the energy spectrum steepens from $k^{-5/3}$ to $k^{-2}$, or even further as the rotation speed is increased.  Morize {\em et al.}\cite{Morize:PF2005} and Staplehurst {\em et al.}\cite{Staplehurst:JFM2008} demonstrated  asymmetry between cyclones and anticyclones.   Moisy {\em et al.}\cite{Moisy:JFM2010} studied decay laws, anisotropy, and cyclone-anticyclone asymmetry in decaying rotating turbulence.  Campagne {\em et al.}\cite{Campagne:PF2014} carried out experiments on rotating turbulence and  demonstrated a dual cascade of kinetic energy.

{A large number of numerical experiments have been performed on rotating turbulence\cite{Morinishi:PF2001, Godeferd:JFM1999, Bardina:JFM1985,Bourouiba:JFM2007}}.   Yang and Domaradzki\cite{Yang:PF2004}, M\"{u}ller and Thiele\cite{Muller:EPL2007}, Mininni {\em et al.}\cite{Mininni:PF2009}, and Biferale {\em et al.}\cite{Biferale:PRX2016} showed that the energy spectrum of rotating turbulence is approximately $k^{-2}$ or $k_\perp^{-2}$.   However, Smith and Lee\cite{Smith:JFM2005} and Sen {\em et al.}\cite{Sen:PRE2012} argued that the energy spectrum is proportional to $k_\perp^{-3}$. Deusebio {\em et al.}\cite{Deusebio:PRE2014} studied how rotating flow transitions from three dimensional to quasi two-dimensional.    Mininni\cite{Mininni:JFM2012} studied the rotating helical turbulence numerically and found that for $k>k_\Omega$, the energy spectrum exponent is $-2.2$, and the helicity spectrum exponent is $-1.8$; { here $k_\Omega$ is the Zeman wavenumber.}  For $k>k_\Omega$, they  observed that the system becomes isotropic, and both energy  and helicity spectra exhibit $k^{-5/3}$ scaling.  

{ Baqui and Davidson\cite{Baqui:PF2015} and Baqui {\em et al.}\cite{Baqui:PF2016} constructed a phenomenological theory of rotating turbulence and argued that the flow is anisotropic with  $E(k_\perp)\sim \epsilon^{2/3} k_\perp^{-5/3} $ and $E(k_\parallel) \sim  \epsilon^{2/3} (L_\parallel/L_\perp)^{2/3} k_\parallel^{-5/3}$, where $L_\parallel$ and $L_\perp$ are respectively the integral length scales parallel and perpendicular to the direction of rotation. }  However, an intriguing aspect of   the results of Baqui and coworkers is the absence of  power law scaling in the inertial range  of $E(k)$. This phenomenon, to the best of our knowledge, stands unexplained, and the  quantitative nature of the spectrum is unreported. {In this paper, we present a model to quantify this spectrum; this is the main result of this paper.} 

{Another feature of rotating turbulence is that it has strong columnar structures. Similar structures are predicted by {the} Taylor--Proudman theorem  in the linear limit (i.e., when the convective term in Eq.~(\ref{eq:u_dim}) is negligible) of rapidly rotating laminar flow\cite{Davidson:book:Turbulence}. Nonlinear interactions among the inertial waves too yield quasi-2D behaviour\cite{Biferale:PRX2016,Bellet:JFM2006,Galtier:PRE2003,Cambon:JFM1997}---a result more relevant to the present study of turbulent rotating flow.}  Note that two-dimensionalization of the flow leads to an inverse cascade of energy (cf. Biferale \textit{et al}.\cite{Biferale:JFM2013,Biferale:PRL2012} and Iyer \textit{et al}.\cite{Iyer:EPJE2015}) that strengthens the columnar structures.   These features go beyond the linear limit.     Using  large-scale simulations on $4096^3$ grid, Biferale {\em et al.}\cite{Biferale:PRX2016} studied in detail the complex structures of rotating turbulence, in particular, the vortical structures and 3D anisotropic fluctuations.

For  {rapidly rotating} flows,  researchers have observed that $u_\perp \gg u_{\parallel}$,  yet $u_{\parallel} \ne 0$. {Here, $u_{\perp}$ and $u_{\parallel}$ are respectively the magnitudes of the velocity components perpendicular and parallel to the rotation axis.} Thus  {rapidly rotating} flow is quasi-2D, not 2D.  As a result, the properties of rotating turbulence differs from 2D hydrodynamic turbulence. For example, the energy spectrum of rotating turbulence at small wavenumbers differs from $k^{-5/3}$, in contrast to $k^{-5/3}$ spectrum observed for 2D hydrodynamic turbulence at low wavenumbers. {As we show later in the paper, the columnar structures are observed in  strongly-rotating flows}. Note that similar features have been observed in magnetohydrodynamic (MHD) turbulence\cite{Sundar:PP2017} and quasi-static MHD turbulence\cite{Verma:ROPP2017,Reddy:PF2014,Favier:PF2010}. Interestingly, there is non-trivial  energy exchange between the perpendicular and parallel components of the velocity \cite{Sundar:PP2017,Verma:ROPP2017,Biferale:PRX2016,Reddy:PF2014,Favier:PF2010,Bellet:JFM2006,Galtier:PRE2003,Cambon:JFM1997}.
There have been various attempts to quantify anisotropy in rotating turbulence.  For example, Delache {\em et al.}\cite{Delache:PF2014} studied scale-by-scale anisotropy using ring spectrum and showed that the flow was nearly isotropic for $k>k_\Omega$, but strongly anisotropy for $k <  k_\Omega$.


Rotation suppresses the energy cascade, hence several researchers have revisited the decay law of total energy.  Thiele and Muller\cite{Thiele:JFM2009} observed that the total energy  $E \sim t^{-1.5}$ for nonrotating flows, but the exponent decreases from 1.5 to $\approx 0.5$ as the rotation speed is increased.   Baqui and Davidson\cite{Baqui:PF2015} also arrived at similar conclusions. Teitelbaum and Mininni\cite{Teitelbaum:PRL2009} however argued that $E \sim t^{-1}$ for nonhelical rotating flows, but $ E \sim t^{-1/3}$ for the helical ones. Here the inverse cascade of energy suppresses the decay of turbulence.

In this paper we investigate the energy spectrum and structures of {rapidly rotating} decaying turbulence. Here we focus on the nonhelical flows (zero kinetic helicity).  We  perform {sufficiently} high resolution spectral simulations that  help us analyze the asymptotic regime. We  go up to $1024^3$ grid resolution and up to $155$ eddy turnover time.  In accordance with the earlier works on rotating turbulence, we observe a strong inverse cascade of energy that strengthens the coherent columnar structures.   The kinetic energy trapped in such structures dissipates very slowly, and hence the Reynolds number remains quite large with slow variation in time.  

However, for {rapidly rotating} decaying turbulence, the energy contents of the intermediate and the small scales are quite small.  We show that the Reynolds number based on the rms speed of these modes is quite small.  Following the arguments similar to Verma {\em et al.}\cite{Verma:arxiv2017}, we show that energy spectrum {rapidly rotating} decaying turbulence is $E(k) \sim k^{-3} \exp(-C (k/k_d)^2)$, where $k_d$ is the enstrophy dissipation wavenumber, and $C$ is a positive real {constant}.   To best of our knowledge, such energy spectrum has not been reported for decaying  {rapidly rotating} turbulence.  

We remark that the above results are for decaying turbulence.  It is {\em a  priori} not obvious that the forced rotating turbulence has similar behaviour as its decaying counterpart.  The energy spectrum could in principle be dependent on the type and nature of external forcing, and on the injection of kinetic helicity. However these topics are beyond the scope of this paper that exclusively deals with the strongly-rotating decaying turbulence. 


The structure of this paper is as follows:  In Section~\ref{sec:Governing_equation}, we briefly review the existing turbulence phenomenologies of rotating turbulence, and then describe our model for {rapidly rotating} decaying turbulence.  Section~\ref{sec:Numerical_Scheme} contains the details of numerical simulations.  In Section~\ref{sec:Taylor}, we describe the properties of the coherent columnar structures in the flow.  In this section, we also discuss the inverse cascade of energy that strengthens such structures.  In Section~\ref{sec:Ek}, we describe the spectra and the fluxes of energy and enstrophy.  We conclude in Section~\ref{sec:conclusion}.


\section{Phenomenology of rotating turbulence}
\label{sec:Governing_equation}

The Navier--Stokes equation of an unforced incompressible fluid  in rotating reference frame is
\begin{eqnarray}
\frac{\partial \bf u}{\partial t} + (\bf u \cdot \nabla) \bf u & = & -{\nabla p} - 2 {\bf \Omega} \times {\bf u} + \nu \nabla^2 {\bf u}, \label{eq:u_dim} \\
\nabla \cdot \bf u & = & 0 \label{eq:inc_dim}, 
\end{eqnarray}
where $\mathbf{u}$ and $p$ are the velocity  and  pressure fields respectively, ${\bf\Omega} =  \Omega \hat{z}$ is the  angular velocity of the rotating reference frame,   $\nu$ is the kinematic viscosity, and $-2 {\bf \Omega} \times {\bf u} $ is the Coriolis acceleration.   The centrifugal acceleration has been absorbed in the pressure gradient term.   We assume the rotation to be along the $z$ direction.  

The ratio of the magnitudes of  $(\bf u \cdot \nabla) \bf u$ and the Coriolis acceleration  is called the Rossby number, i.e.,
\be 
\mathrm{Ro} = \frac{U_0}{\Omega L_0},
\label{eq:Ro}
\ee
{ where $U_0$ and $L_0$ are the large  velocity and length scales respectively. }  Coriolis force drives the perpendicular component of the velocity field, ${\bf u}_\perp = u_x \hat{{ x}} + u_y \hat{{ y}}$; this is one of the reasons why the  {rapidly rotating} flows tend to be quasi-2D with  $u_\perp \gg u_\parallel$ {($u_{\parallel}$ is the magnitude of the velocity component along $z$ direction)}\cite{Chakraborty:EPJB2010,Chakraborty:EPL2007,Baroud:PF2003,Canuto:PF1997_rotation,Bartello:JFM1994,Hossain:PF1994}.    In the present paper we focus on {rapidly rotating} decaying flows, i.e. for very small $\mathrm{Ro}$. Note that the Reynolds number $\mathrm{Re} = U_0 L_0/\nu$. 

Owing to the quasi-2D nature of the flow, the Kolmogorov's phenomenology for the 3D hydrodynamic turbulence does not apply to rotating turbulence. Researchers have proposed different models for rotating turbulence; some of these models are described below. Zeman\cite{Zeman:PF1994} argued that the Coriolis force dominates the nonlinear advection term for $k<k_\Omega$, and vice versa for $k>k_\Omega$, where $k_\Omega$ is called {\em Zeman scale}.  The Kolmogorov's $k^{-5/3}$ spectrum is expected to hold for $k> k_\Omega$.  Zeman\cite{Zeman:PF1994} derived an expression for $k_\Omega$ by equating the advection term and the Coriolis acceleration at $k=k_\Omega$, i.e.,  $
k_\Omega u_{k_\Omega}^2 \sim \Omega u_{k_\Omega}$.   The Kolmogorov's phenomenology for hydrodynamic turbulence yields $u_k \sim \epsilon^{1/3} k^{-1/3}$, substitution of which in $
k_\Omega u_{k_\Omega}^2 \sim \Omega u_{k_\Omega}$ yields the Zeman wavenumber as
\be
k_\Omega = \sqrt{\frac{\Omega^3}{\epsilon}}.
\label{eq:k_omega}
\ee
For  $k<k_\Omega$, Zeman\cite{Zeman:PF1994} employed scaling arguments and argued that $E(k) \sim \epsilon^{2/5} \Omega^{4/5} k^{-11/5}$.  This phenomenology resembles the energy spectrum of stably-stratified turbulence with buoyancy\cite{Obukhov:DANS1959,Verma:NJP2017}.  

In another phenomenology, Zhou\cite{Zhou:PF1995} modelled rotating turbulence in a spirit similar to the Iroshnikov's\cite{Iroshnikov:SA1964} and Kraichnan's\cite{Kraichnan:PF1965MHD} model of magnetohydrodynamic turbulence and derived that $E(k) \sim (\epsilon \Omega)^{1/2}   k^{-2}$; here the relevant time scale was taken to be $\Omega^{-1}$.   { Zhou\cite{Zhou:PF1995} generalized the above phenomenology to include nonlinear time scale, and obtained dual scaling with $k^{-2}$ spectrum for $k \ll k_\Omega$, and $k^{-5/3}$ spectrum for $k \gg k_\Omega$.}  Smith and Waleffe\cite{Smith:PF1999} equated the nonlinear advection term with the Coriolis force and obtained $E(k) \sim \Omega^{2} k_\perp^{-3}$, where $k_\perp$ is the horizontal wavenumber.  Using perturbative approach, Chakraborty\cite{Chakraborty:EPL2007} showed that $E(k) \sim k^{-2.87}$ for weakly rotating turbulent systems.  In latter part of this section we show that for  rapidly rotating turbulence ($\mathrm{Ro} \ll 1$), the energy spectrum tends to be of the form $k^{-3}\exp(- C (k/k_d)^2)$, { where $k_d$ is the enstrophy dissipation wavenumber, and $C$ is a positive real  constant.}  It is important to note that the  Coriolis force does not do any work on the fluid.


Researchers have attempted to verify the above phenomenologies using experiments and numerical simulations.   Some of the these works have been described in Section~\ref{sec:intro}, and we do not repeat them here.  In the following discussion we derive a model of $E(k)$  based on the flux variation with $k$.

For rapid rotation, the flow tends to be strongly quasi two-dimensional, i.e., $u_z \ll u_\perp$.  {The Taylor--Proudman theorem\cite{Davidson:book:Turbulence} predicts such structures in the linear regime. But in the nonlinear regime, structure formation is  due to the inverse cascade of energy (see Section~\ref{sec:intro}). Note that in the linear regime,  $\mathrm{Re} =0$, and hence $\Pi(k) =0$.  However, in the nonlinear regime, as will be shown in our numerical simulations,  large-scale vortices are formed due to strong nonlinear effects.  }  Therefore we study  ${\bf u}_\perp$ by taking a horizontal cross section of the flow profile.  It is best to relate the 2D-sectional field with 2D hydrodynamic theory of Kraichnan\cite{Kraichnan:PF1967_2D}.  It is important however to keep in mind that the rotating flow is more complex due to the $u_z$ component that couples with ${\bf u}_\perp$.   {  External forcing is absent in decaying turbulence, but  small wavenumber modes supply energy to the large wavenumber modes. Therefore, decaying 2D turbulence and rapidly rotating turbulence exhibit forward enstrophy cascade\cite{Clercx:AMR2009}.}



In Fourier space,  one-dimensional energy spectrum is defined as $E(k) = \sum_{k -1 < k^{\prime} \leq k} \frac{1}{2} |{\bf {u}}({\bf k^\prime})|^2$, whose evolution equation is given by
\begin{equation}
\frac{\partial }{\partial t} E(k,t)= -\frac{\partial }{\partial k}  \Pi(k,t)  - 2 \nu k^2 E(k,t), \label{eq:energy_time}
\end{equation}
where $\Pi(k,t) $ is the energy flux emanating from a wavenumber sphere of radius $k$ at time $t$.  For a steady or a quasi-steady state, $\partial E(k)/\partial t \approx 0$, hence
\begin{equation}
 \frac{d}{dk} \Pi(k) = -2 \nu k^2 E(k). \label{eq:dPibydk}
\end{equation}

The energy flux and the spectrum, $\Pi(k)$ and $E(k)$, are two unknown functions whose solution cannot be obtained from a single equation, Equation~(\ref{eq:dPibydk}).  For 3D hydrodynamic turbulent flows, Pao\cite{Pao:PF1965} assumed that $E(k)/\Pi(k)$ is independent of $\nu$, and that it depends only on $\epsilon$ and $k$. Under these assumptions, we  obtain the following solution for the above:
\begin{eqnarray}
E(k) & = & K_\mathrm{Ko} \epsilon^{2/3} k^{-5/3} \exp{\left(- \frac{3}{2} K_\mathrm{Ko}  (k/k_\eta)^{4/3}\right)}, \label{eq:Pao_Ek} \\
\Pi(k) & =  & \epsilon \exp{\left(- \frac{3}{2} K_\mathrm{Ko}   (k/k_\eta)^{4/3}\right)}, \label{eq:Pao_Pik}
\end{eqnarray}
where $\epsilon$ is the energy dissipation rate, $K_\mathrm{Ko}$ is Kolmogorov's constant, and $k_\eta = (\epsilon/\nu^3)^{1/4}$ is Kolmogorov's  wavenumber. { \citet{Verma:arxiv2017} generalized the aforementioned Pao's phenomenology to laminar regime ($\mathrm{Re} \lessapprox 1$).}


{Since the  rapidly rotating flows tend to be quasi-2D, it is important to briefly describe the phenomenology of 2D hydrodynamic turbulence.   Kraichnan\cite{Kraichnan:PF1967_2D} had reported that the two-dimensional hydrodynamic turbulence has dual spectrum---the energy exhibits inverse cascade for $k < k_f$, while the enstrophy $E^{(2D)}_\omega = \int d{\bf r}  \omega^2/2 = \int d{\bf r} |\nabla \times {\bf u}|^2/2$ exhibits forward cascade for $k > k_f$, where  $k_f$ is the forcing wavenumber.  Kraichnan\cite{Kraichnan:PF1967_2D} showed that for $k< k_f$,
\bea
E^{(2D)}(k) & = & K'_\mathrm{Ko} [\Pi^{(2D)}(k)]^{2/3} k^{-5/3} \label{eq:Ek_k_less_kf}
\eea
with $\Pi^{(2D)}(k) = \mathrm{const} <0$. However, for $k>k_f$, 
 \bea
 E_\omega^{(2D)}(k) & = &K_\omega [\Pi^{(2D)}_\omega(k)]^{2/3} k^{-1} \label{eq:Ek_k_greater_kf}
 \eea
 with $ \Pi^{(2D)}_\omega(k) = \mathrm{const} >0$.  Here $\Pi^{(2D)}(k)$ and   $\Pi^{(2D)}_\omega(k)$ are  the energy and enstrophy fluxes respectively,  $K'_\mathrm{Ko}$ is Kolmogorov's constant for 2D,  $K_\omega$ is the proportionality constant for the constant enstrophy flux regime.}

{In the subsequent discussion, we will show that for strongly-rotating turbulence, the intermediate and small-scale structures contain very small amount of energy.  Also, since the flow is quasi-2D, we study the enstrophy and energy of a horizontal cross section perpendicular to the rotation axis.  The  evolution equation for the enstrophy spectrum is
\be
\frac{\partial }{\partial t} E^{(2D)}_\omega(k,t)= -\frac{\partial }{\partial k}  \Pi^{(2D)}_\omega(k,t)  - 2 \nu k^2 E^{(2D)}_\omega(k,t), \label{eq:enstrophy_time}
\ee 
where $\Pi^{(2D)}_\omega(k,t)$ is the enstrophy flux for the 2D flow of a horizontal cross-section.   For a steady or a quasi-steady state, $\partial E^{(2D)}_\omega(k,t)/\partial t \approx 0$, hence
 \begin{equation}
 \frac{d}{dk} \Pi^{(2D)}_\omega(k) = -2 \nu k^2 E^{(2D)}_\omega(k).\label{eq:dPi_omea_bydk}
\end{equation}
For the  intermediate and small scales,  following Pao\cite{Pao:PF1965}, we assume that  $E^{(2D)}_\omega(k)/\Pi^{(2D)}_\omega(k)$ is independent of $\nu$, and  it  depends only on the enstrophy dissipation rate, $\epsilon_\omega$, and $k$.  Under this ansatz,   $E^{(2D)}_\omega(k)$ and $\Pi^{(2D)}_\omega(k)$ are given by
\begin{eqnarray}
E^{(2D)}_\omega(k) & = & C \epsilon_\omega^{2/3} k^{-1} \exp{\left(-  C   (k/k_d)^{2}\right)}, \label{eq:Ek_omega_k1}\\
\Pi_{\omega}^{(2D)}(k) & =  & \epsilon_\omega \exp{\left(- C   (k/k_d)^{2}\right)}, \label{eq:Pik_omega_k1}
  \label{eq:Pao}
\end{eqnarray}
where 
\be
k_d = \frac{\epsilon_\omega^{1/6}}{\sqrt{\nu}}
\label{eq:k_d}
\ee
is the enstrophy dissipation wavenumber, and $\epsilon_\omega$ is the enstrophy dissipation rate. }

For strongly-rotating turbulence discussed in this paper, $\mathrm{Re} \gg 1$, and the flow is quasi-2D.  In this paper we show some subtle differences between the 2D hydrodynamic turbulence and the rotating turbulence.  We show that  Eqs.~(\ref{eq:Ek_omega_k1}, \ref{eq:Pik_omega_k1})  match quite well with our numerical results apart from prefactors. These results will be discussed in Section~\ref{sec:Ek}.

We also remark that in the linear regime { where nonlinearity is absent,}
\be
\frac{\partial }{\partial t} E^{(2D)}_\omega(k,t)=   - 2 \nu k^2 E^{(2D)}_\omega(k,t),\label{eq:enstrophy_time}
\ee 
which has solution of the form\cite{Kraichnan:PF1967_2D},
\be
  E^{(2D)}_\omega(k,t)=     E^{(2D)}_\omega(k,0) \exp(- 2 \nu k^2 t).
	\label{eq:enstrophy_time_evolution}  
  \ee
  More importantly, the enstrophy flux must be zero for this case.  { This predictions is not applicable to our numerical results due to the presence of the nonlinear {interactions}, however weak, and nonzero enstrophy flux. }

In the following section, we compare the phenomenology developed in this section with numerical results.

\section{Numerical Simulation}
\label{sec:Numerical_Scheme}
To investigate the dynamics of  decaying rotating non-helical turbulent flow, we  perform direct numerical simulation (DNS) using {pseudo-spectral} code Tarang\cite{Chatterjee:JPDC2018,Verma:Pramana2013tarang}.  We  solve Equations~(\ref{eq:u_dim}) and (\ref{eq:inc_dim}) in a 3D periodic and cubic box of size $(2\pi)^3$. We  use the fourth-order Runge--Kutta scheme for time stepping,   {Courant--Friedrichs--Lewy} (CFL) condition  to obtain optimal time step ($\Delta t$), and $2/3$-rule for  dealiasing. {We take $\nu = 10^{-3}$ and $\Omega = 16$.  Note that these parameters are nondimensional, in the sense that $ \nu/(U'L') \rightarrow \nu$ and $\Omega L'/U' \rightarrow \Omega$ with $U', L'$ as the characteristic velocity and length scale of the system in proper dimension (e.g. $L'$  in meters).  We remark that $U' \ne  U_0$, where $U_0$ is the nondimensional  rms velocity of the system.   For our simulations we  employ grid-resolutions of $512^3$ and $1024^3$. We observe that these grids yield  similar results, thus we verify grid-independence of our results.} For all our runs, $k_{\rm max}\eta > 1$, where $\eta$ is the Kolmogorov's length, and $k_{\rm max}$ is the highest wavenumber represented by the grid points.  Hence all our simulations are well-resolved.   Note that in the above, $k_\mathrm{max} = N/2$ with $N$ as the grid size. 

\begin{figure}[htbp]
\begin{center}
\includegraphics[scale=1.0]{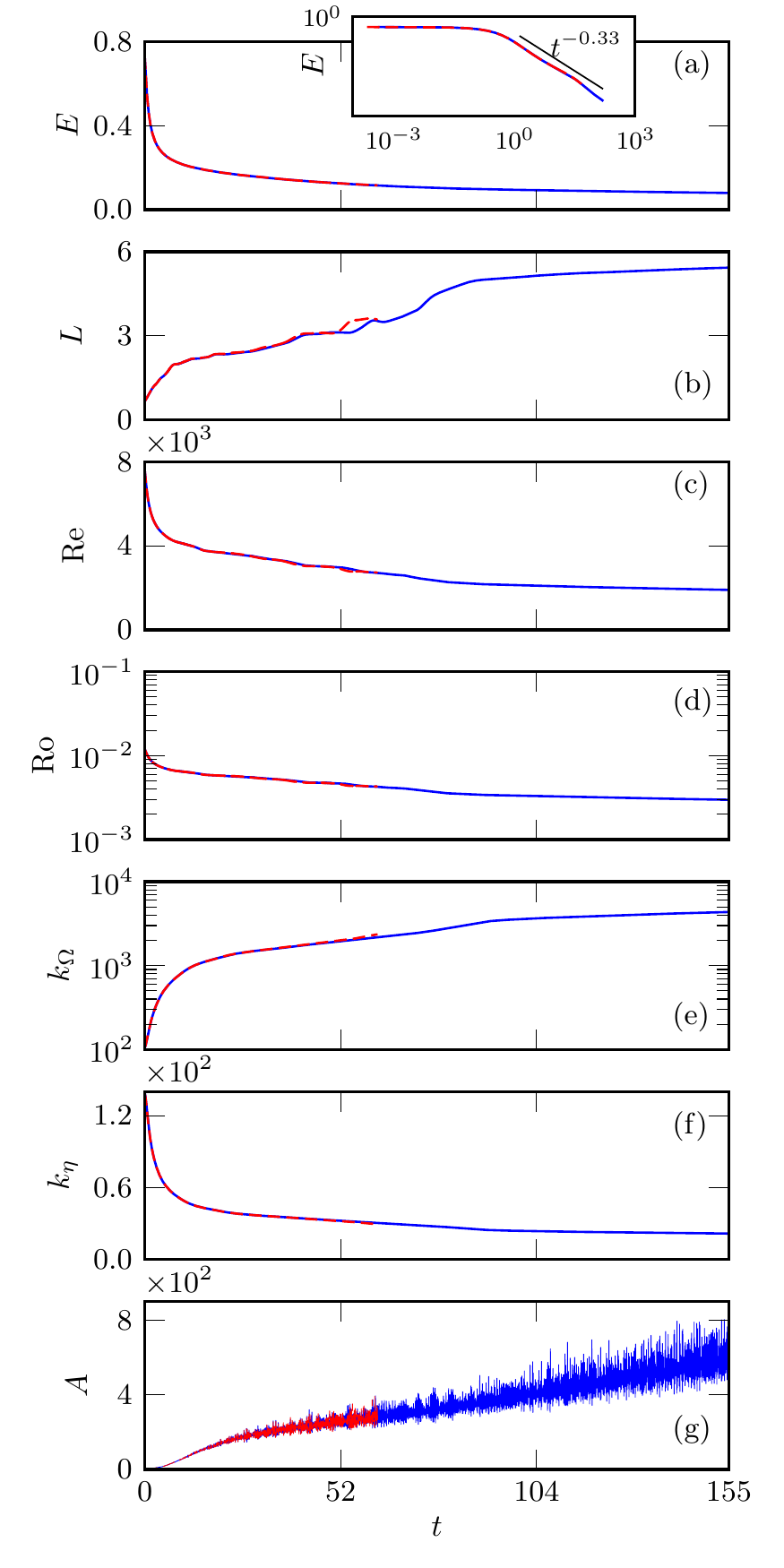}
\end{center}
\setlength{\abovecaptionskip}{0pt}
\caption{For {$1024^3$ (red dashed) and }$512^3$ (blue) grids simulation with $\Omega=16$, variation of  characteristic system parameters with time $t$: (a) the total energy $E(t)$, (b) the integral length scale, $L$, (c) the Reynolds number, ${\rm Re}$, (d) the Rossby number, ${\rm Ro}$, (e) the Zeman wavenumber, $k_{\Omega}$,  (f) the Kolmogorov dissipation wavenumber, $k_{\eta}$, and (g) anisotropy of the system $A$. The inset in subfigure (a) is log-log plot showing $E(t) \sim t^{-0.33}$.}
\label{fig:1}
\end{figure}

First we generate a fully-developed hydrodynamic turbulence ($\Omega = 0$) with $\nu = 10^{-3}$ on  $512^3$ grid with random forcing  in the  wavenumber band of $(11,12)$.   Our forcing is such that it supplies a constant energy and zero kinetic helicity to the flow, which is achieved by a force field
\be 
{\bf f(k)} = \alpha_k {\bf u(k)},~~\mathrm{where}~\alpha_k = \frac{\epsilon}{N_f |{\bf u(k)}|^2}.
\ee
Here $\epsilon$ is the energy supply rate, and $N_f$ is the total number of wavenumber modes in the forcing band where ${\bf f}$ is employed\cite{Carati:JoT2006}.
 
Now we use the steady-state data of {the three-dimensional homogeneous isotropic fully-developed hydrodynamic turbulence} as an initial condition for the simulation of our {rapidly rotating} turbulence. {Note that  such initial conditions have been widely used in earlier simulations of decaying and rotating turbulence\cite{Baqui:PF2015,Delache:PF2014,Yoshimatsu:JFM2011}. We employ  $\Omega=16$.  Note that the (nondimensional) frequency associated with the nonlinearity is $O(1)$, hence our simulation with $\Omega=16$  is reasonably fast rotating.   } We carry out simulation of the rotating flow with the same forcing as hydrodynamic simulation for 6 nondimensional time units after which the forcing is  turned off.  We set $t=0$ here. The strongly-rotating and decaying simulation starts at this stage, and it is carried out till $t=t_{\rm final} = 155$.  We also carry out  simulation of the rotating flow on $1024^3$ grid with the same set of parameters and initial conditions.  Since $1024^3$ simulation is much more expensive, we end this simulation at $t=49$.

In Figure~\ref{fig:1} we show temporal evolution of  various quantities for $512^3$ {(blue) and $1024^3$ (red)} grids.  Figure~\ref{fig:1}(a) exhibits the evolution of total energy, $E = \int d{\bf r} u^2/2$, that  decays from 0.8 to approximately 0.08 following a power law  $E \sim t^{-0.33}$.  This result is in general agreement with those of Thiele and Muller\cite{Thiele:JFM2009} and Baqui and Davidson\cite{Baqui:PF2016}.  In Figure~\ref{fig:1}(b) we plot the integral length scale defined by
\be 
L = 2\pi \frac{\int dk k^{-1} E(k)}{\int dk E(k)}.
\ee
We observe that beyond $t=100$, $L \approx 5.43$, which is close to the box size ($2\pi$), thus signalling formation of large scale structures.  As shown in Figure~\ref{fig:1}(c), the {Reynolds} number $\mathrm{Re} = U_0 L_0/\nu$ is in the range of 2000 to 3000. {For the computation of $\mathrm{Re}$ and $\mathrm{Ro}$ we employ $L_0=2\pi$ (box size), and $ U_0$ as the rms speed, which is given by $\left(2 \int_{0}^{k_{\mathrm{max}}} E(k) dk\right)^{1/2}$}.


\setlength{\tabcolsep}{15.0pt}
\begin{table}[htpb]
\begin{ruledtabular}
\caption{Parameters of the direct numerical simulations (DNS): List of total energy $E$; integral length scale $L$; Reynolds number $\mathrm{Re}$; Rossby number $\rm{Ro}$; Zeman wavenumber $k_{\Omega}$; Kolmogorov dissipation wavenumber $k_{\eta}$ ; and anisotropy ratio $A$ at  $t=49, 148$ for $512^3$ grids, and at $t=49$ for $1024^3$ grids.}
\label{table:numerical_parameters}
{\begin{tabular}{c c c c} 
Parameters    &   {$t = 49$ }    &  $t=148$   & $t=49$  \\ 
              &   {$(N = 512^3)$}      &  $(N = 512^3)$  & $(N = 1024^3)$ \\ \hline
$E$           &   {$ 0.13 $ }     &  $0.08$    & $0.13$  \\ 
$L$           &   {$ 3.11 $ }    &  $5.40$    & $3.10$  \\
$\rm{Re}$     &   {$ 3003 $ }    &  $1918$    & {2989}  \\
$\rm{Ro} $    &   {$ 0.005 $}   &  $0.003$   & $0.005$  \\
$k_{\Omega}$  &   {$ 1888 $ }    &  $4300$    & $1900$  \\
$k_{\eta}$    &   {$ 32 $ }      &  $21$      & $32$  \\
$A$           &   {$ 228 $ }     &  $596$     & $241$  \\ 
\end{tabular}}
\end{ruledtabular}
\end{table}

We compute the Rossby number $\mathrm{Ro}$ using Equation~(\ref{eq:Ro}) and plot its temporal variation in Figure~\ref{fig:1}(d). The figure shows that $\mathrm{Ro}$ varies from $10^{-2}$ to $3 \times 10^{-3}$.  Thus, Rossby number at large times is quite small for our simulations. Hence, the Coriolis force dominates the nonlinear term of Equation~(\ref{eq:u_dim}).  We remark that some researchers\cite{Baqui:PF2016} report $\mathrm{Ro}$ based on the initial velocity of decaying turbulence.  However our definition is based on the instantaneous velocity, thus $\mathrm{Ro}$ of Figure~\ref{fig:1}(d) is that of instantaneous flow. 

We also compute the Zeman wavenumber $k_\Omega$ using Eq.~(\ref{eq:k_omega}),  and the Kolmogorov's wavenumber $k_\eta$ using $(\epsilon/\nu^3)^{1/4}$, and plot them in Figure~\ref{fig:1}(e,f) respectively.  We observe that in the asymptotic regime, $k_\Omega  \gg 1$ indicating dominance of Coriolis force. {
We observe that  $k_{\Omega} > k_\mathrm{max}$, where $k_\mathrm{max} = N/2$ with $N$ as the grid size. Hence the  above estimate of $k_\Omega$ appears to be quite ambiguous. Note that the derivation of $k_\Omega$ using Eq.~(\ref{eq:k_omega}) assumes that  for large $k$'s, the turbulence is isotropic and  $E(k) \sim k^{-5/3}$, which is not the case for our simulations of strongly-rotating turbulence (to be discussed in Sec.~\ref{sec:Ek}).  We show that strongly rotating turbulence makes the flow quasi-2D.  Hence, strictly speaking,  Eq.~(\ref{eq:k_omega}) cannot be employed to compute $k_\Omega$.}

  In Figure~\ref{fig:1}(g) we  plot the anisotropy parameter
\be
A = \frac{E_\perp}{2E_\parallel}
\ee
as a function of time.  Here, $E_{\perp} = E_{x} + E_{y}$, and $E_{\parallel} = E_{z} $ [with $E_x = \int (u_x^2/2) d{\bf r}$, $E_y = \int(u_y^2/2) d{\bf r}$, and $E_z = \int (u_z^2/2) d{\bf r}$]. We observe that $A \gg 1$ indicating quasi-2D nature of the flow. We have also tabulated the values of $E$, $L$, $\mathrm{Re}$, $\mathrm{Ro}$, $k_{\Omega}$, $k_{\eta}$, and $A$ in Table~\ref{table:numerical_parameters} at  $t=49, 148$ for $512^3$ grid simulations, and at $t=49$ for $1024^3$ grid simulation respectively. 

In the next section we show that the strongly-rotating flow is dominated by the columnar structures.


\section{Columnar structures and associated Fourier modes}
\label{sec:Taylor}
In the earlier section we  showed that the global parameters like the integral length scale and the anisotropy parameter indicate presence of large scale structures.  In this {Section} we describe these structures along with their associated Fourier modes.
\begin{figure*}[htbp]
\begin{center}
\includegraphics[scale=1.0]{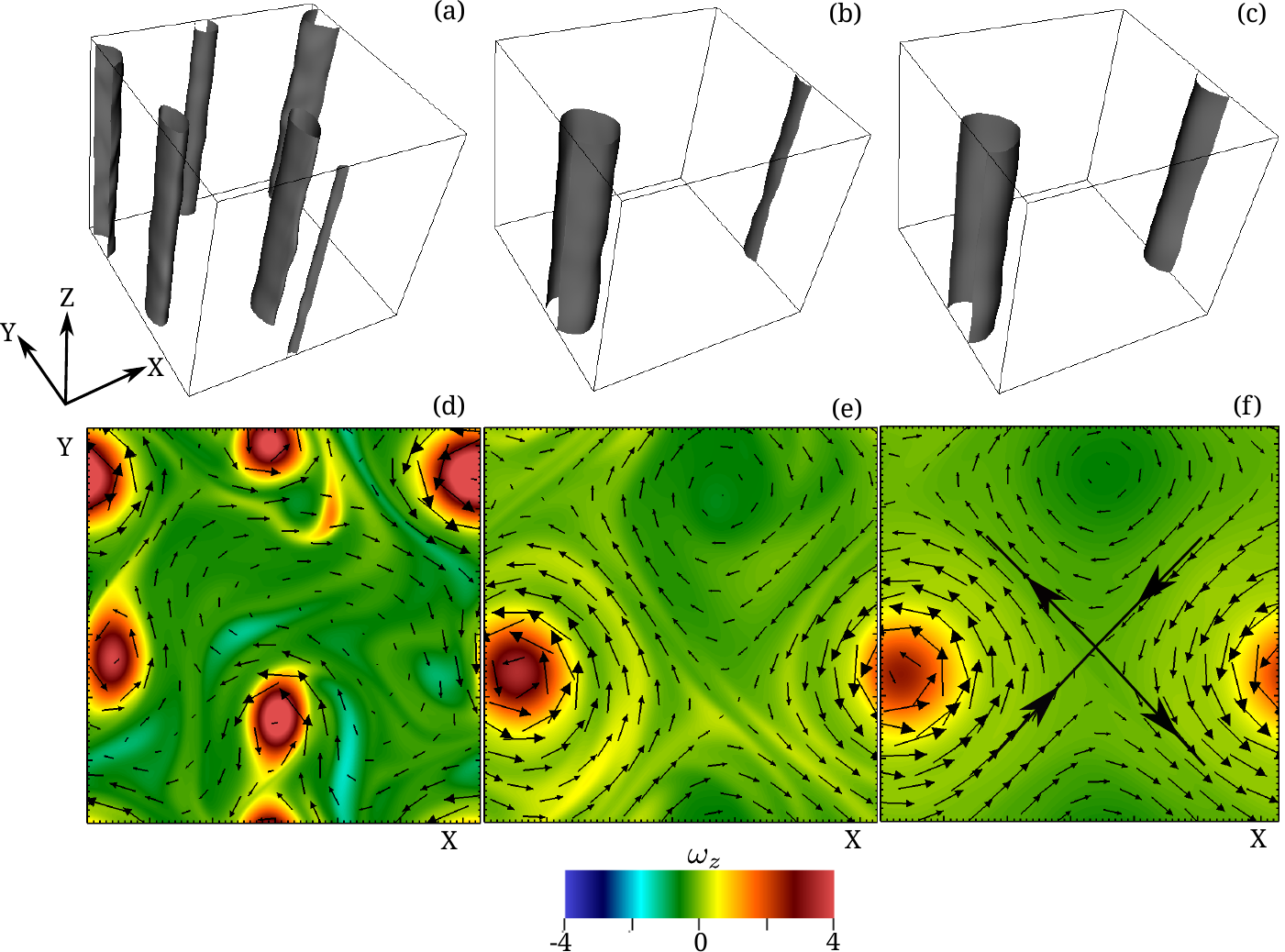}
\end{center}
\setlength{\abovecaptionskip}{0pt}
\caption{For the {rapidly rotating} decaying turbulence on $512^3$ grid: The top panel exhibits  the isosurfaces of the magnitude of vorticity $|\boldsymbol{\omega}|$ at (a) $t=49$, (b) $t=98$, and (c) $t= 148$. The bottom panel shows velocity vector plot superposed with the density plot for $\omega_z$ for the horizontal cross section at $z = \pi$ at  (d) $t = 49$, (e) $t=98$, and (f) $t=148$. {Note that Fig.~(f) contains only one vortex that appears to be split in two due to the periodic boundary conditions.} }
\label{fig:2}
\end{figure*}


\begin{table}
\begin{ruledtabular}
\caption{For {rapidly rotating} turbulence on $512^3$ grid at $t=4,49,148$,  the energy contents of the dominant modes as percentage of the total energy. Note that ${\bf u}(-k_x,-k_y,0)$, not listed in the table, has the same energy contents as that of ${\bf u}(k_x,k_y,0)$. In the table, $E_\mathrm{mode}=|u({\bf k})|^2/2$.  { For the modes  $(0,1,0)$ and $(1,0,0)$, $E_{\mathrm{mode}}/E$ increases with time indicating strengthening of the vortical structures with time.}}
\label{table:total_energy}
{\begin{tabular}{c  c c c} 
Mode &   {$E_{\mathrm{mode}}/E$  (\%)}   &  { $E_{\mathrm{mode}}/E$  (\%)}    &$E_{\mathrm{mode}}/E$  (\%) \\
($k_x$, $k_y$, $k_z$) & {$t = 4$}  &  {$t = 49$}  & $t = 148$ \\ [0.1mm] \hline 
{$(0,1,0)$} & {$0.04$} & {$3.45$} & {$22.12$} \\ [0.1mm]
{$(1,0,0)$}  & {$1.21$} & {$7.34$} & {$20.79$} \\ [0.1mm]
               {$(1,1,0)$}  & {$0.60$} & {$6.20$} &                {$1.88$} \\ [0.1mm]
               {$(-1,1,0)$} & {$0.32$} & {$1.75$} &                {$1.66$}\\ [0.1mm]
               {$(2,1,0)$}  & {$0.65$} & {$1.60$} &                {$0.47$}  \\[0.1mm] 
{$(2,-1,0)$} & {$1.08$} & {$2.17$} & {$0.44$} \\ [0.1mm]
               {$(-2,1,0)$} & {$1.08$} & {$2.17$} &                {$0.44$}  \\ [0.1mm]
               {$(1,2,0)$}  & {$0.74$} & {$1.02$} &                { $0.41$} \\ [0.1mm]
               {$(1,-2,0)$} & {$0.50$} & {$1.92$} &                {$0.40$} \\ [0.1mm]
               {$(2,-2,0)$} & {$0.38$} & {$1.23$} &                {$0.11$}\\ [0.1mm]
{$(-3,0,0)$} & {$0.68$} & {$0.36$} & {$0.10$} \\ [0.1mm]
               {$(2,2,0)$}  & {$0.52$} & {$0.96$} &                {$0.10$} \\ [0.1mm]
               {$(-1,-3,0)$} & {$1.70$}& {$0.54$} &                {$0.05$} \\ [0.1mm]
               {$(3,1,0)$}   & {$0.29$}& {$1.04$} &                {$0.05$} \\ [0.1mm]
               {$(-3,1,0)$}  & {$2.80$}& {$0.19$} &                { $0.05$} \\ [0.1mm]
               {$(3,-2,0)$}  & {$0.45$}& {$0.19$} &                {$0.02$} \\ [0.1mm]
               {$(-2,-3,0)$} & {$1.67$}& {$0.26$} &                {$0.02$}  \\ [0.1mm]
               {$(-3,-2,0)$} & {$0.89$}& {$0.04$} &                {$0.02$}  \\ [0.1mm] \\
\hline
Total \%:                    & {$13.084$} & {$32.43$}  & {$49.13$}\\ [0.1mm]
\hline
\end{tabular}}
\end{ruledtabular}
\end{table}


We investigate the large scale structures by studying the flow profile in real space.  In Figure~\ref{fig:2}(a,b,c), we exhibit the contour plots of the magnitude of the vorticity field,~$|\omega|$, at $t=49$, 98, and 148.  These figures demonstrate existence of strong vortical  structures.  To decipher the flow profiles of these columns, we take horizontal cross section of the flows at $z=\pi$, and present the density plots of $\omega_z$ superposed with the vector plot of ${\bf u}_\perp = u_x \hat{x} + u_y \hat{y}$.  At $t=49$ we observe four cyclonic vortices that have anticlockwise sense of rotation (see Figure~\ref{fig:2}(d)).  Note the periodicity of the box. Subsequently these vortices merge and form a single cyclonic vortex, as shown in Figure~\ref{fig:2}(c,f).  In a periodic box these vortices reside on a lattice along with weak anti-cyclonic vortices. { These features are quite similar to those in 2D hydrodynamic turbulence~\cite{Clercx:AMR2009}. }

The cyclonic and anti-cyclonic vortices are separated by a  {\em saddle}, which is symbolised by a cross in Figure~\ref{fig:2}(f). Morize {\em et al.}\cite{Morize:PF2005} and Staplehurst {\em et al.}\cite{Staplehurst:JFM2008} observed  cyclonic/anti-cyclonic asymmetry in experiments, while  van Bokhoven {\em et al.}\cite{VanBokhoven:JoT2008} quantified the asymmetry between cyclonic and anti-cyclonic by studying skewness of vertical vorticity.  Smith and Lee\cite{Smith:JFM2005} argued that the cyclonic/anti-cyclonic asymmetry arise due to nonlinear interactions near resonance.

{ We remark that the size of the asymptotic (at large time) flow structures described above are proportional to the box size.  We demonstrate this feature in Appendix~\ref{appA} by simulating rotating turbulence in two boxes of sizes $(2\pi)^3$ and $(4\pi)^3$.  These results show that the Fourier modes and their interactions are independent of the box size.} 

When we compare the flow structures of Figure~\ref{fig:2}(e,f) with those found in two-dimensional Hamiltonian dynamics, the cross  and the centres of the vortices of Figure~\ref{fig:2}(f) correspond to the  {\em saddle} and the  {\em centres}.  This similarity is due to the divergence-free condition of the velocity field that yields
\be
 \nabla \cdot {\bf u}_\perp = 0 \implies 
 \frac{\partial \dot{x}}{\partial x} + \frac{\partial \dot{y}}{\partial y} = 0,
 \ee
 which is analogous to the equation for the conservation of phase space area of a two-dimensional Hamiltonian system\cite{Strogatz:book}.  Note that $u_z \ll u_\perp$, hence we treat our system as two-dimensional for the above discussion.

The emergence of large scale structures in the flow can be quantified using the energy contents of  small wavenumber Fourier modes, which are listed in Table~\ref{table:total_energy}.  { Evidently, in the asymptotic regime ($t=148$), the  Fourier mode ${\bf k} = (k_x, k_y, k_z) = (1,0,0)$ and (0,1,0) are the most dominant modes, with the other strong Fourier modes being $(1,1,0)$ and $(-1,1,0)$. Note that we do not list the energies of $-{\bf k}$ modes because ${\bf u(-k) = u^*(k)}$.  Hence, the energies $E(-1,0,0) = E(1,0,0)$ and $E(0,-1,0) = E(0,1,0)$.  When we add the energies of $(\pm 1,0,0)$ and $(0,\pm 1,0)$, we observe that they contain approximately 80\% of the total energy.  These modes form a strong set of 2D vortices, as discussed in Appendix~\ref{appB} and Figure~\ref{fig:8}.  The nonlinear dynamics of these modes is very interesting, and it may shed light on the cyclone-anticyclone antisymmetry in the presence of rotation. But these discussions are beyond the scope of this paper.} 

{ At $t=148$, the sum the energies of the 18 dominant modes listed in Table~\ref{table:total_energy} and those of their complex conjugate partners is approximately 98\%.  These modes lie within the sphere of radius 4. Hence,  modes in the intermediate and small scales contain very small amount of energy. This result has a strong consequence on the energy and enstrophy spectra of the strongly-rotating turbulence, which will be discussed in Sec.~\ref{sec:Ek}. }

  In the following discussion we argue why the small wavenumber modes become strong in rapidly rotating turbulence.   The strong vortical structures of the flow indicate quasi-2D nature of the flow.  This observation is reinforced by the fact that $A =  E_\perp / (2 E_\parallel) \gg 1$ (see Figure~\ref{fig:1}(g)).   The flow become quasi-2D because of the Coriolis force that is active in the perpendicular plane ($x,y$), as well as due to the inverse cascade of the kinetic energy from  small scales (large $k$)  to  large scales (small $k$). This is in contrast to the emergence of the Taylor columns in the linear limit, as predicted by {the} Taylor--Proudman theorem;  the energy transfer is absent all together in the linear limit.

\begin{figure}[htbp]
\begin{center}
\includegraphics[scale=1.0]{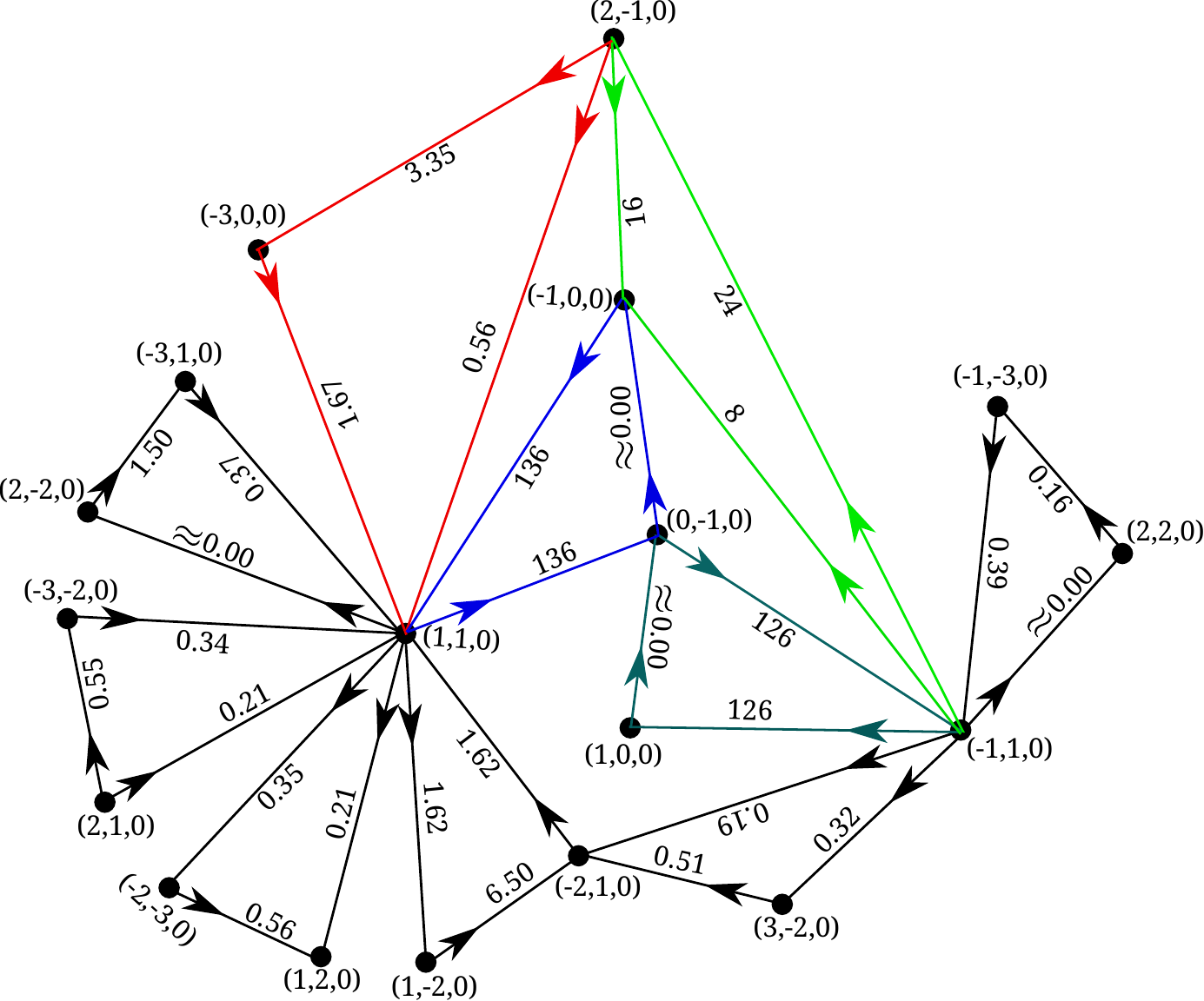}
\end{center}
\setlength{\abovecaptionskip}{0pt}
\caption{ {For the rapidly rotating decaying  simulation on $512^3$ grid, the energy transfers $S({\bf k'|p|q})$ for some of the dominant triads averaged over five eddy turnover time from  $t=148$ to $152$.   The numbers above the arrows are $S({\bf k'|p|q}) \times 10^{5}$ for convenient description.  The most dominant energy transfers are in the triad $[(-1,0,0), (1,-1,0), (0,-1,0)]$. }   }
\label{fig:3}
\end{figure}

  Let us now quantify the above observations using the mode-to-mode energy transfers and the energy flux.  First we describe the energy transfers among the large scale Fourier modes.  Dar {\em et al.}\cite{Dar:PD2001} and Verma\cite{Verma:PR2004} showed that for an interacting triad of fluid flow (${\bf k',p,q}$) that satisfies the relation ${\bf k' + p +q}=0$, the rate of energy transfer from mode ${\bf u(p)}$ to mode ${\bf u(k')}$ with mode ${\bf u(q)}$ acting as a mediator is
\begin{equation}
 S(\mathbf{k}'| \mathbf{p}| \mathbf{q}) = -\mathrm{Im}[ ({\bf k' \cdot  u(q)}) ({\bf u(p) \cdot  u(k')})].\label{eq:Skpq}
 \end{equation}
To investigate the growth of the large scale structures, we study the energy transfers among the small wavenumber modes listed in Table~\ref{table:total_energy}.    Figure~\ref{fig:3} exhibits some of the dominant interacting triads involving small $k$ Fourier modes.   Note that a complete graph with $N$ modes would contain approximately $N(N-1)/2$ edges, which is quite large for $N\sim 10$.  Hence we show only some of the representative energy transfers. The numbers above the arrows represent the energy transfers.  { The most dominant energy transfer is in the triad $[{\bf k}' = (1,1,0), {\bf p} =(-1,0,0), {\bf q} = (0,-1,0)]$ with the mode ${\bf u}(-1,0,0)$ supplying approximately $ 136\times 10^{-5}$ units of energy to the mode ${\bf u}(1,1,0)$.  We observe that the dominant energy transfers are $(-1,0,0) \rightarrow (1,1,0)$ $\rightarrow (0,-1,0)$ $\rightarrow (-1,1,0)$ $\rightarrow (-1,0,0)$.  Thus, the four modes $(1,0,0), (0,1,0), (1,1,0), (-1,1,0)$ play a critical role in the energy transfers in rotating turbulence. These issues will be studied in more detail in future. Figure~\ref{fig:3}  also exhibits other dominant energy transfers, but these transfers are order of magnitude smaller than those discussed above. }


After the aforementioned discussion on the large scale structures and their associated Fourier modes, in the next section we present the energy and enstrophy spectra and fluxes of our system.

\section{Energy and enstrophy fluxes and spectra}
\label{sec:Ek}

The energy spectrum and flux provide valuable information about the flow.  In this section we compute these quantities and study their features.  

\begin{figure}[htbp]
\begin{center}
\includegraphics[scale=1.0]{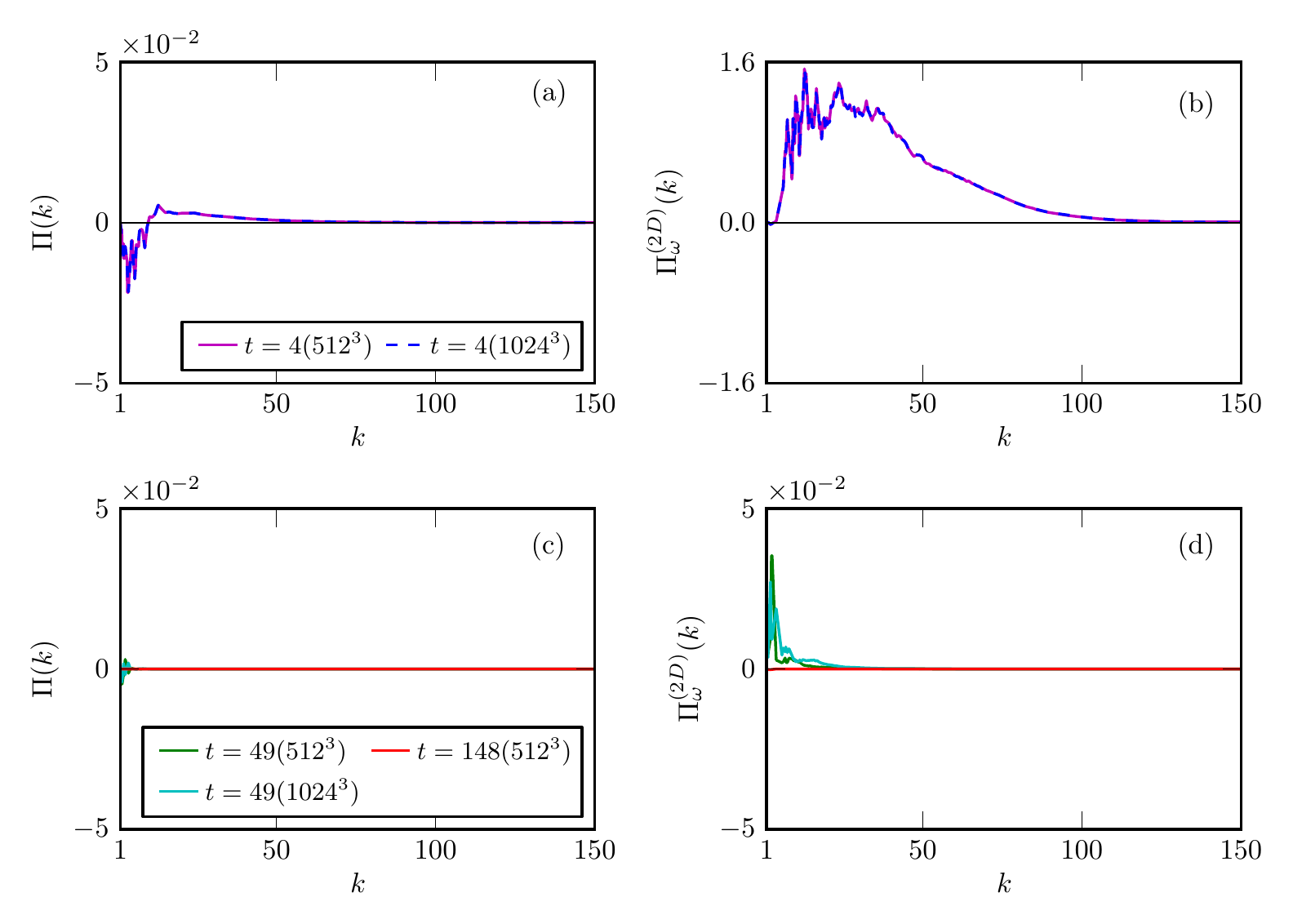}
\end{center}
\setlength{\abovecaptionskip}{0pt}
\caption{{Kinetic energy flux $\Pi(k)$ and enstrophy flux  $\Pi_{\omega}^{(2D)}(k)$ for {rapidly rotating} decaying turbulence.  (a)  $\Pi(k)$ of 3D velocity field for  $512^3$ (magenta) and $1024^3$ (blue dashed curve) grids at  $t = 4$.   (b)    $\Pi_{\omega}^{(2D)}(k)$ for the 2D cross section at $z = \pi$. Figure (b) has the same color convention as (a). (c)  $\Pi(k)$  at $t = 49$ (green), and $t = 148$ (red) for the grid resolutions of $512^3$ and at $t = 49$ (cyan) for $1024^3$ grids.   (d) $\Pi_{\omega}^{(2D)}(k)$ for  2D cross section at $z = \pi$. Figure (d) has the same color convention as (c).  Note that $\Pi_{\omega}^{(2D)}(k)$ is significant, but $\Pi(k)$ is negligible. }}
\label{fig:4}
\end{figure}

We compute the energy flux at $t=4$, $49$ and $148$ using the following formula\cite{Verma:PR2004}:
\be
\Pi(k_0) =\sum_{k'>k_0} \sum_{p \le k_0}  S({\bf k'|p|q}),
 \label{eq:Pi_k_compute}
\ee
where $k_0$ is the radius of the wavenumber sphere from whom the flux is being computed,  $S({\bf k'|p|q})$ is defined in Equation~(\ref{eq:Skpq}), and ${\bf k'+p+q=0}$.  We compute the energy flux at $t = 4,49,148$ using the $512^3$ and $1024^3$ grid data.  These results are plotted in Figure~\ref{fig:4}(a,c) for $t = 4$ (magenta), {$t = 49$ (green)} and $148$ (red) of $512^3$ grid simulation, and for $t = 4$ (blue dashed) and $49$ (cyan)  of $1024^3$ grid simulation. We observe a significant inverse cascade of kinetic energy at  early stages.   As time progresses, the kinetic energy flux becomes weaker and gets concentrated in the wavenumber band $k\in[1,8]$. This feature is compatible with the strongly vortical quasi-2D structure of the flow.    Ours is a decaying simulation, so  we expect the inverse energy cascade regime to be narrower, and the forward enstrophy cascade regime to be effective for a larger wavenumber range.   For this reason, it is more appropriate to study the enstrophy spectrum and flux, for which we employ the data obtained from horizontal cross sections  at $z=\pi/2$, $\pi$, and $3\pi/2$.


We compute the enstrophy flux of the 2D velocity field ${\bf u}_\perp$ at the planes $z=\pi/2,\pi$, and $3\pi/2$ using 
\be
\Pi^{(2D)}_\omega (k_0) =  \sum_{k>k_0} \sum_{p \le k_0}  S^{\omega\omega} ({\bf k'|p|q}),
 \label{eq:Pi_enstrophy_k_compute}
\ee
where
\be
S^{\omega\omega} ({\bf k'|p|q}) = -\mathrm{Im}[ ({\bf k \cdot  u_\perp(q)}) \omega_z({\bf p}) \omega_z({\bf k'}) ]
\ee
represents the enstrophy transfer from  mode $\omega_z({\bf p})$ to mode $\omega_z({\bf k})$ with mode ${\bf u_\perp(q)}$ acting as a mediator.  Note that ${\bf u_\perp} = u_x \hat{x} + u_y \hat{y}$ and $\omega_z ({\bf k}) = [i {\bf k} \times {\bf u(k)}]_z$.  For the cross section at $z = \pi$, Figure~\ref{fig:4}(b,d) illustrates the plots  $\Pi_{\omega}^{(2D)}(k)$ vs. $k$  at $t = 4$, {$49$}, $148$ for the grid resolution of $512^3$, and at $t = 4$, $49$ for the grid resolution of $1024^3$.  { For $512^3$ and $1024^3$ grids  at the same time, the energy and enstrophy fluxes are equal, which is consistent with the fact that our results are grid-independent.}

{ The enstrophy flux is positive definite, but it is not constant in a significant wavenumber band, in contrast to 2D hydrodynamic turbulence for which $\Pi^{(2D)}_\omega (k)$ is constant in the inertial range and then it decreases after $k=k_d$\cite{Clercx:AMR2009}.  The steepening of  $\Pi^{(2D)}_\omega (k)$ in strongly-rotating turbulence is  due to the viscous effects, as in Equation~(\ref{eq:Pik_omega_k1}), and due to energy transfer from ${\bf u}_\perp$ to $u_z$, analogous  to that in quasi-static MHD\cite{Verma:ROPP2017,Reddy:PP2014}; this is in contrast to constant $\Pi^{(2D)}_\omega (k)$ in hydrodynamic two-dimensional turbulence for $k>k_f$.}

\begin{figure*}[htbp]
\begin{center}
\includegraphics[scale=0.85]{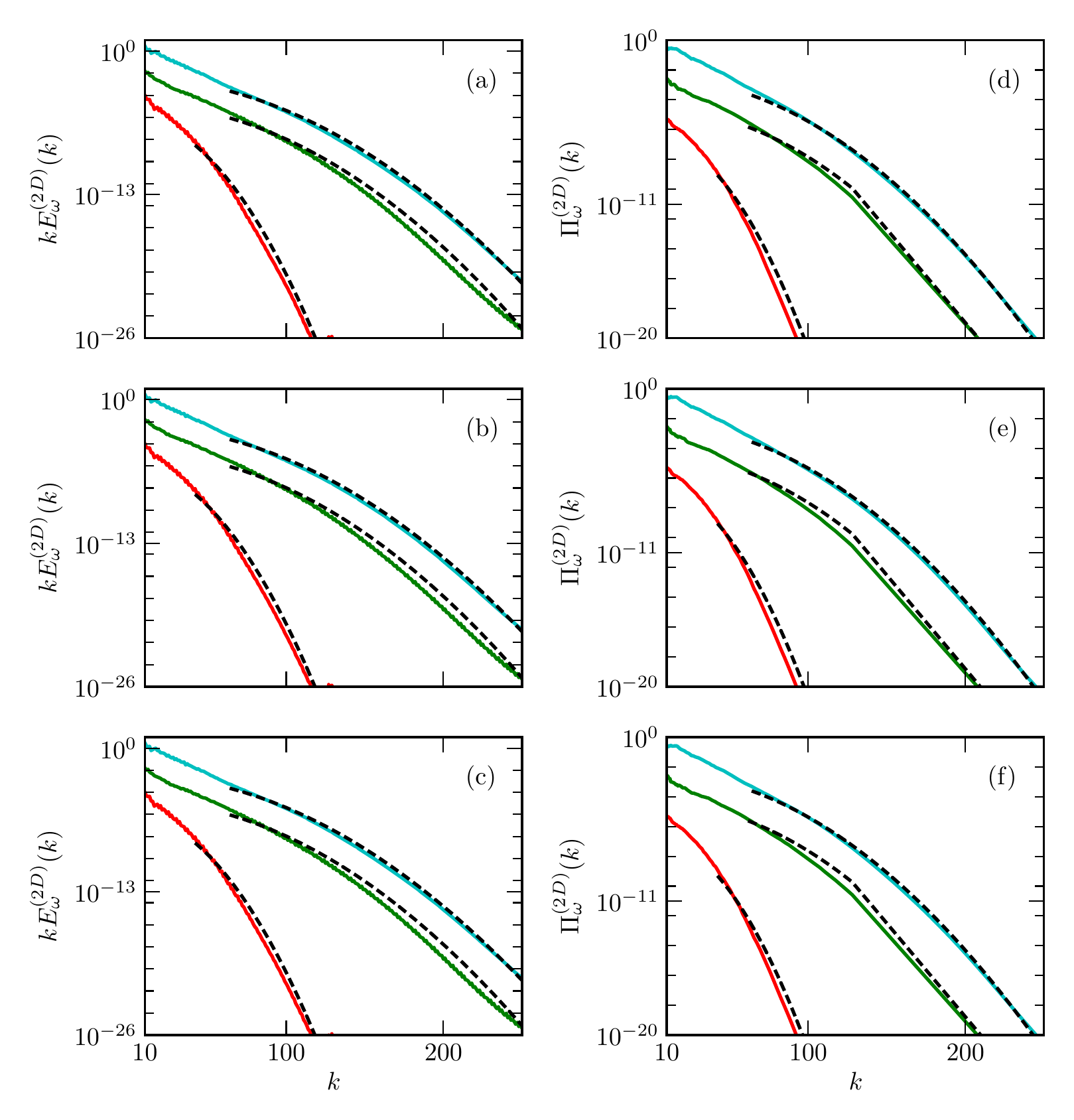}
\end{center}
\setlength{\abovecaptionskip}{0pt}
\caption{For the {rapidly rotating} decaying simulation using ${\bf u}_\perp$  at $z=\pi/2,\pi,3\pi/2$.    Two-dimensional enstrophy spectrum $kE^{(2D)}_\omega(k)$ for (a) $z=\pi/2$, (b) $z=\pi$,  (c) $z=3\pi/2$ at {$t = 49$ (green) and} $t=148$ (red) for $512^3$ grid simulation, and at $t=49$ (cyan) for the $1024^3$ grid. The best fit curves using Equation~(\ref{eq:Ek_omega_k1}) with parameters of Table~\ref{table:parameters} are shown  as black dashed curves.  (d,e,f) The corresponding plots of the 2D enstrophy flux $\Pi^{(2D)}_{\omega}(k)$  following the same color convection as above.  The best fit curves however are of the form of Equation~(\ref{eq:Pik_omega_k1}).  { Since $E^{(2D)}_\omega(k)$ and $\Pi^{(2D)}_{\omega}(k)$  of $512^3$ and $1024^3$ grids almost overlap on each other, we multiply  $E^{(2D)}_\omega(k)$ and $\Pi^{(2D)}_{\omega}(k)$ of $1024^3$ grid with a factor $1/100$ to differentiate the two plots.}}
\label{fig:5}
\end{figure*}

Now let us focus on the time frames $t = 49,\,148$ when the coherent columnar structures are well developed and strong. For these times, in Figure~\ref{fig:5}, we plot {$kE_{\omega}^{(2D)}(k)$} and $\Pi_{\omega}^{(2D)}(k)$ vs. $k$ in semi-log scale for $z=\pi/2$, $\pi$, and $3\pi/2$ planes.   In rotating turbulence,  $\Pi^{(2D)}_\omega (k)$ starts to decrease at small $k$ itself because the enstrophy dissipation wavenumber, $k_d$, is quite small (see Table III).  We compute the enstrophy dissipation wavenumber $k_d$ using Equation~(\ref{eq:k_d}).   {It is difficult to estimate $\epsilon_{\omega}$ of Equation (\ref{eq:Ek_omega_k1}) because we do not have a band of wavenumbers where $\Pi_{\omega}^{(2D)}(k)$ is constant.  In this paper, we compute $\epsilon_{\omega}$ using Equation (\ref{eq:Pik_omega_k1}) by identifying  the neck of the  wavenumber range from where $\exp(-C(k/k_d)^2)$ spectrum starts.  If the wavenumber at the neck is $k_*$, then using   Equation (\ref{eq:Pik_omega_k1}), 
\begin{equation}
\epsilon_{\omega} = \Pi_{\omega}^{(2D)}(k_*)\exp(C(k_*/k_d)^2).
\end{equation}
Incidentally we observe that $k_*$ is approximately twice of Kolmogorov's wavenumber, $k_{\eta}$.  We compute $C$ using linear regression analysis. The values of $k_d$, $\epsilon_{\omega}$, and $C$ are tabulated in Table~\ref{table:parameters}.  In Figure~\ref{fig:5},  we plot the best fits to the enstrophy flux as dashed black curves (see  Equation (\ref{eq:Pik_omega_k1})).  We observe that the above equation describes the numerical data very well. Also note that  $\Pi^{(2D)}_{\omega}(k)$  of $512^3$ and $1024^3$ grids almost overlap on each other.  Hence, to contrast the two plots, we multiply  $E(k)$ and $\Pi^{(2D)}_{\omega}(k)$ of $1024^3$ grid with a factor $1/100$ to differentiate the two plots.}

Motivated by the above observations, we model the enstrophy spectrum $E^{(2D)}_\omega(k)$ using Equation~(\ref{eq:Ek_omega_k1}).  As shown in Figure~\ref{fig:5}, we observe the numerical results to be in very good agreement with the model of Equation~(\ref{eq:Ek_omega_k1}).  The red {and the green curves} in the Figure~\ref{fig:5}  represent $kE_\omega^{(2D)}(k)$  and $\Pi_\omega^{(2D)}(k)$ for $512^3$ resolution at {$t=49$ and $t=148$} {respectively}, while the cyan curve represents the corresponding plots for $1024^3$ resolution at $t=49$. Also note that as expected, $E^{(2D)}_\omega(k)$ and $\Pi^{(2D)}_\omega(k)$ decrease with time due to their decaying nature. Thus we claim that the enstrophy spectrum and  flux for the {rapidly rotating} decaying turbulence are described by Equations (\ref{eq:Ek_omega_k1}, \ref{eq:Pik_omega_k1}) respectively.  

\setlength{\tabcolsep}{6pt}
\begin{table}
\begin{ruledtabular}
\caption{For the {rapidly rotating} decaying turbulence, we take the instantaneous 2D velocity field on the horizontal cross sections at $z=\pi/2$, $\pi$, and $3\pi/2$.  List of the  enstrophy dissipation rate, $\epsilon_\omega$,   the enstrophy dissipation wavenumber, $k_d$, and  constant $C$.  These parameters are listed in the table at time $t=49$  {for $512^3$ and $1024^3$}, and $t = 148$ for $512^3$. }
\label{table:parameters}
{\begin{tabular}{c c c c c c c  c c}

\text{Grid} &$t$ & $ z $ &  $k_d$   & $\epsilon_\omega \times 10^{5}$    &  $C \times 10^{2}$ \\[1 mm] 
\hline
{$512^3$}  & {$49$}  & {$\pi/2$}   & {$6.0$}     & {$4.8$}  & {$(2.73 \pm 0.04) $}\\
{$512^3$}  & {$49$}  & {$\pi$}     & {$6.0$}     & {$4.9$}  & {$(2.76 \pm 0.04)$} \\
{$512^3$}  & {$49$}  & {$3\pi/2$}  & {$6.0$}     & {$4.9$}  & {$(2.74 \pm 0.04)$} \\
\hline
{$1024^3$}  & {$49$}  & {$\pi/2$}   & {$6.6$}     & {$8.4$}  & {$(3.00 \pm 0.02) $}\\
{$1024^3$}  & {$49$}  & {$\pi$}     & {$6.6$}     & {$8.7$}  & {$(3.03 \pm 0.02)$} \\
{$1024^3$}  & {$49$}  & {$3\pi/2$}  & {$6.6$}     & {$8.2$}  & {$(2.98 \pm 0.02)$} \\ 
\hline
{$512^3$}   & {$148$} & {$\pi/2$}   & {$2.7$}     & {$0.04$}  & {$(2.45 \pm 0.05)$} \\
{$512^3$}   & {$148$} & {$\pi$}     & {$2.7$}     & {$0.04$}  & {$(2.51 \pm 0.05)$} \\
{$512^3$}   & {$148$} & {$3\pi/2$}  & {$2.7$}     & {$0.04$}  & {$(2.48 \pm 0.05) $} \\
\hline
\end{tabular}}
\end{ruledtabular}
\end{table}

\begin{figure}[htbp]
\begin{center}
\includegraphics[scale=0.85]{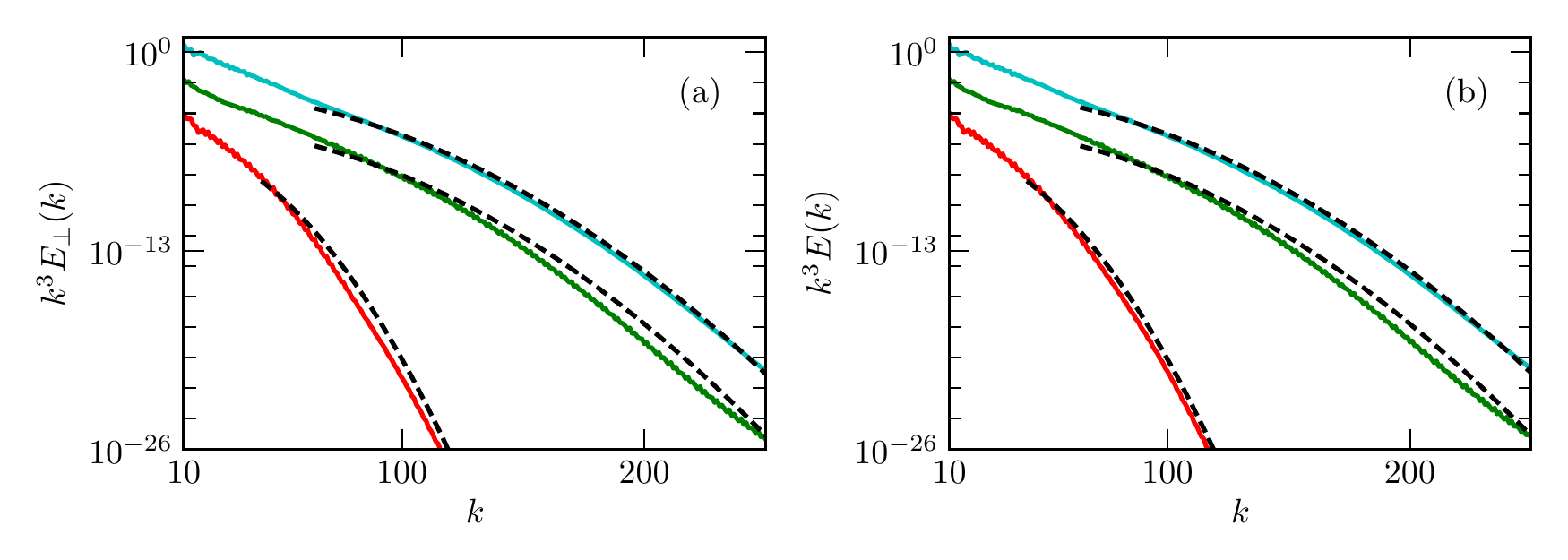}
\end{center}
\setlength{\abovecaptionskip}{0pt}
\caption{ {For the {rapidly rotating} decaying turbulence: (a) Normalized 3D energy spectrum $k^3 E_\perp(k)$ vs. $k$ at {$t=49$ (green)},$148$ (red) for $512^3$ grid simulation, and at $t=49$ (cyan) for  $1024^3$ grid. The best fit curves following Equation~(\ref{eq:Ek_fit}) with $\epsilon_\omega$ and $C$ of Table~\ref{table:parameters} are plotted as black dashed curves.  (b) The corresponding plots of $k^3 E(k)$ where $E(k) = E_\perp(k)+E_\parallel(k)$.  Since $E(k)$  of $512^3$ and $1024^3$ grids almost overlap on each other, we multiply  $E(k)$ of $1024^3$ grid with a factor $1/100$ to differentiate the two plots. } }
\label{fig:6}
\end{figure}

We now model the energy spectrum for the full flow.  Using ${\boldsymbol{\omega}} = \nabla \times {\bf u}$ and Equation (\ref{eq:Ek_omega_k1}), we deduce that
\bea
E^{(2D)}(k) & = &  \frac{1}{k^2} E^{(2D)}_\omega (k)  \nonumber \\ 
& =  & C \epsilon_\omega^{2/3} k^{-3} \exp\left(-C   (k/k_d)^{2}\right).
\label{eq:E_perp_k}
\eea
For the  cube (3D flow), 
\be
E_\perp(k) = E_x(k) + E_y(k).
\ee
Though $E_\perp(k)$ is not strictly equal to $\sum_{k_z} E^{(2D)}(k)$, we observe that $E_\perp(k)$  has similar scaling as $E^{(2D)}(k)$, i.e.,
\be
E_\perp(k) =C^{\prime\prime} \epsilon_\omega^{2/3} k^{-3} \exp\left(-C   (k/k_d)^{2}\right).
\ee
where $C^{\prime\prime}$ is another constant. {This is evident from Figure~\ref{fig:6}(a), in which we plot $k^3 E_\perp(k)$ vs.~$k$ and the best fit curves as dashed black lines, $C \epsilon_\omega^{2/3} \exp\left(-C   (k/k_d)^{2}\right)$, at $t=49$, $148$ for $512^3$ and $t = 49$ for $1024^3$ grid resolution.} In the figure, the red {and green curves} represent the numerical spectra for the $512^3$ resolution (at $t = 49, 148$), while the cyan curve is for the $1024^3$ resolution (at $t=49$). We observe that $C^{\prime\prime} \approx C$ because  $E_z(k) \ll E_\perp(k)$.   In the plots of Figure 6 too, at $t=49$, the energy spectra of $512^3$ and $1024^3$ grids almost overlap with each other. Therefore, we multiply $E(k)$ of $1024^3$ grids with 1/100 to contrast the two plots.

In {rapidly rotating} flows, the velocity component along $z$ direction is strongly suppressed, hence $E_z(k) \ll E_\perp(k)$.  This observation is supported by the plot of the anisotropy parameter of Figure~\ref{fig:1}(g).   
Hence
\bea
E(k) & = & E_x(k) + E_y(k)+E_z(k) \approx E_\perp(k)  \nonumber \\ 
& & \approx  C \epsilon_\omega^{2/3} k^{-3} \exp\left(-C   (k/k_d)^{2}\right).
\label{eq:Ek_fit}
\eea
In Figure~\ref{fig:6}(b) we plot $k^3 E(k)$ following the same convention as Figure~\ref{fig:6}(a).  We observe that the Equation~(\ref{eq:Ek_fit}) fits with the numerical data quite well with the same $C$ as of Figure~\ref{fig:5}.  Thus, we claim that $E(k) \approx   C \epsilon_\omega^{2/3} k^{-3} \exp\left(-C   (k/k_d)^{2}\right)$ for the {rapidly rotating} decaying turbulence.


As described in Section~\ref{sec:intro}, for rapidly rotating turbulence, many researchers have reported dual spectrum for $E(k)$ with larger wavenumbers exhibiting Kolmogorov's spectrum.  There is no unanimity on the spectral index for low wavenumber modes ($k<k_\Omega$) with  researchers \cite{Biferale:PRX2016,Baqui:PF2016,Baqui:PF2015,Sen:PRE2012,Mininni:JFM2012,Mininni:PF2009,Muller:EPL2007,Smith:JFM2005,Yang:PF2004} predicting the spectral   exponents as $-2$, $-3$, $-5/3$.
In fact, Morize {\em et al.}\cite{Morize:PF2005} argue that the spectrum steepens with the increase of rotation speed.  For details refer to Section~\ref{sec:intro}.   We did attempt to fit a power law $(E(k) \sim k^{-\alpha})$ with our data and  observed that the spectral indices are $-8$ or lower for very small range of low wavenumbers. These observations yield stronger confidence in the  model of Equations~(\ref{eq:Ek_omega_k1}, \ref{eq:Pik_omega_k1}) that evidently spans over a much longer range of wavenumbers.  { We also remark that our model for $E(k)$ (Equation (\ref{eq:Ek_fit})) performs better than those of Pao and Kraichnan, which are described by Equations (\ref{eq:Pao_Ek}) and Equation (\ref{eq:enstrophy_time_evolution}) respectively. Note that the enstrophy and energy fluxes in Kraichnan's model are zero. See Appendix C for details.}

We conclude in the next section. 
 
\section{Conclusions and discussion}
\label{sec:conclusion}

In this paper we have performed numerical simulation of {rapidly rotating}  decaying turbulence.    In the asymptotic regime, the  Rossby number of the flow is quite small, and the Reynolds number is quite large.  Strong rotation causes quasi two-dimensionalization of the flow and formation of large coherent columnar structures.  Most of the kinetic energy is concentrated in these structures, and the fluctuations in the inertial and the dissipative ranges have very small amount of energy leading to small Reynolds number for these fluctuations.  Thus, {rapidly rotating} decaying turbulence has strong flow structures embedded in a sea of fluctuations of small magnitudes. 

We have shown that the columnar structures are formed due to strong inverse cascade of energy.    The vortex columns are quasi-2D with $u_z \ll u_\perp$, so we study the 2D energy and enstrophy spectra and fluxes of ${\bf u}_\perp$ for various horizontal cross sections.  We observe that the kinetic energy flux is quite small for $k > 8$, but the enstrophy flux, $\Pi^{(2D)}_\omega$ is significant in this regime. Further, we deduce the expression for the enstrophy spectrum and the enstrophy flux for such flows as Equations~(\ref{eq:Ek_omega_k1}, \ref{eq:Pik_omega_k1}) respectively. Since $u_z \ll u_\perp$, $E(k)  \approx E_\omega^{(2D)}(k)/k^2$, thus  $E(k) \sim C \epsilon_\omega^{2/3} k^{-3} \exp\left(-C   (k/k_d)^{2}\right)$.

The anisotropy induced by intense rotation has strong similarities with those induced by strong mean magnetic field in MHD and quasi-static (QS) MHD turbulence\cite{Verma:ROPP2017,Sundar:PP2017}.  The flows in the rotating turbulence, as well as in MHD and QS MHD turbulence with strong mean magnetic field, are quasi-2D with  $u_\perp \gg u_\parallel$.  Also, the spectra of QS MHD and rotating turbulence for large $k$ have exponential behaviour\cite{Verma:ROPP2017}.  Further quantification of anisotropy in these systems would yield interesting insights. 

The results presented in the paper indicates that the strong rotation induces strong vortical structures.  Note that our simulations have very small Rossby number in comparison to the earlier simulations.  Weak rotation (Rossby numbers of order 1) is likely to yield energy spectrum with power laws $(E(k) \sim k^{-\alpha})$.  { It will be interesting to make a comprehensive study of variations of turbulence properties with the variation of Ro.  However such study is beyond the scope of this paper.}  Note that forced rotating turbulence  yields different energy spectrum, as shown by earlier researchers\cite{Biferale:PRX2016,Baqui:PF2016,Baqui:PF2015,Sen:PRE2012,Mininni:JFM2012,Mininni:PF2009,Muller:EPL2007,Smith:JFM2005,Yang:PF2004}.  Also, kinetic helicity plays a major role in rotating flows. Thus, a detailed comparative study of the present work on strongly-rotating turbulence with those on forced and helical rotating turbulence is in order; these studies will be carried out in future.  

{ Many astrophysical, geophysical, and engineering flows involve strong rotation.  For example\cite{Dormy:book:Dynamo}, Earth's outer core has $\mathrm{Ro} \approx 10^{-6}$, and solar convection has $\mathrm{Ro} \approx 10^{-2}$. The results discussed in this paper may be relevant to  such systems.  Note however that such systems typically involve thermal convection or magnetoconvection, hence they are more complex than what is discussed here.}



\appendix
\section{{Finite-size effects}}
\label{appA}

\begin{figure}[htbp]
\begin{center}
\includegraphics[scale=1.0]{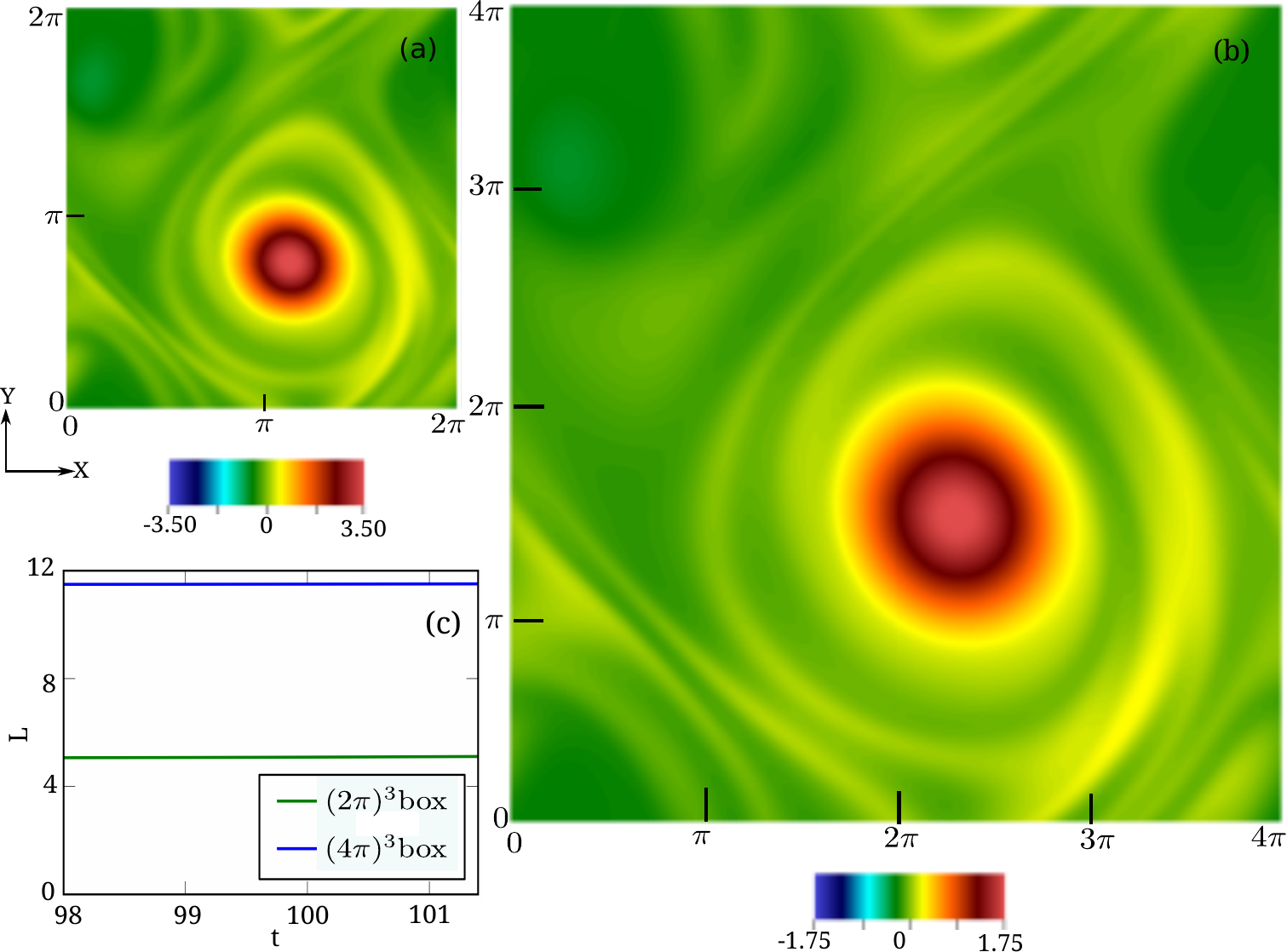}
\end{center}
\setlength{\abovecaptionskip}{0pt}
\caption{{For rapidly rotating turbulence, the density plots of $\omega_{z}$ at $z=\pi$ in boxes of sizes $(2\pi)^3$ (a) and $(4\pi)^3$ (b).  For initial condition, we take the $512^2$ data at $t=49$ of Fig.~\ref{fig:2}. Clearly, the size of the vortex in $(4\pi)^3$ box is twice that of $(2\pi)^3$ box.  (c) Plots of the time series of integral length scales $L$ in $(2\pi)^3$ (green curve) and $(4\pi)^3$ (blue curve).  Here $L_{4\pi} \approx 2 L_{2\pi}$.}}
\label{fig:7}
\end{figure}

{To verify the finite size scaling, we performed numerical simulations of rotating turbulence in two boxes of sizes $(2\pi)^3$ and $(4\pi)^3$.  For the initial condition of these runs, we took our $512^3$ simulation data at $t=98$.  In the $(4\pi)^3$ box, ${\bf u(x)}$ outside the $(2\pi)^3$ box (central region) was set to zero  at $t=0$. The simulation was carried out till $t_f=104$.  In Figure~\ref{fig:7}(a)-(b), we exhibit density plots of the vorticity field at $z=\pi$, i.e., $\omega_z(x,y,z=\pi)$,  at $t=100$.  We observe that the size of the vortex in the $(4\pi)^3$ box is twice compared to that in $(2\pi)^3$ box. Hence, the large-scale vortex is indeed due to nonlinear effects, and it is independent of the box size.  We also compute the integral length scales $L$ for the two boxes, and observe them to be approximately 5.1 and 11.5 respectively.  See Fig.~\ref{fig:7}(c) for an illustration.  Clearly, the integral length scale for $(4\pi)^3$ box is approximately double of that of $(2\pi)^3$ box.  Thornber~\cite{Thornber:PF2016} studied the impact of domain size and statical error in decaying turbulence.  Our preliminary studies on strongly-rotating turbulence appear to show that its results are somewhat immune to system size.  That is, the Fourier modes and their interactions are independent of the box size. }


\section{{Fourier modes $(1,0,0)$ and $(0,1,0)$ in rotating turbulence}}
\label{appB}

{ In Sec.~\ref{sec:Taylor} we showed that $(1,0,0)$ and $(0,1,0)$ are the most dominant Fourier modes of strongly-rotating turbulence.  In this section we discuss the 2D flow pattern when the system has only $(1,0,0)$ and $(0,1,0)$ Fourier modes. Here we choose the velocity field as
\begin{equation}
{\bf u}(x,y) = \hat{x} \sin y + \hat{y} \sin x.
\end{equation}
In Figure~\ref{fig:8}, we exhibit the velocity field  along with the density plot of $\omega_z$ for the above field.  The flow pattern is quite similar to that of Figure~\ref{fig:2}(f), except that Figure~\ref{fig:8}(a) shows cyclone-anticyclone symmetry, but Figure~\ref{fig:2}(f) is asymmetric in cyclone-anticyclone pattern. Clearly, the cyclone-anticyclone asymmetry arises due to rotation.  

In Appendix~\ref{appA} we described the results of numerical simulations of rapidly rotating turbulence.  In Figure~\ref{fig:8}(b) we exhibit the time series of energies of Fourier modes $(1,0,0)$, $(0,1,0)$, $(1,1,0)$, and $(-1,1,0)$, along with the total energy $E_T$. Clearly, the modes $(1,0,0)$, $(0,1,0)$ dominate other modes.


\begin{figure}[htbp]
\begin{center}
\includegraphics[scale=1.0]{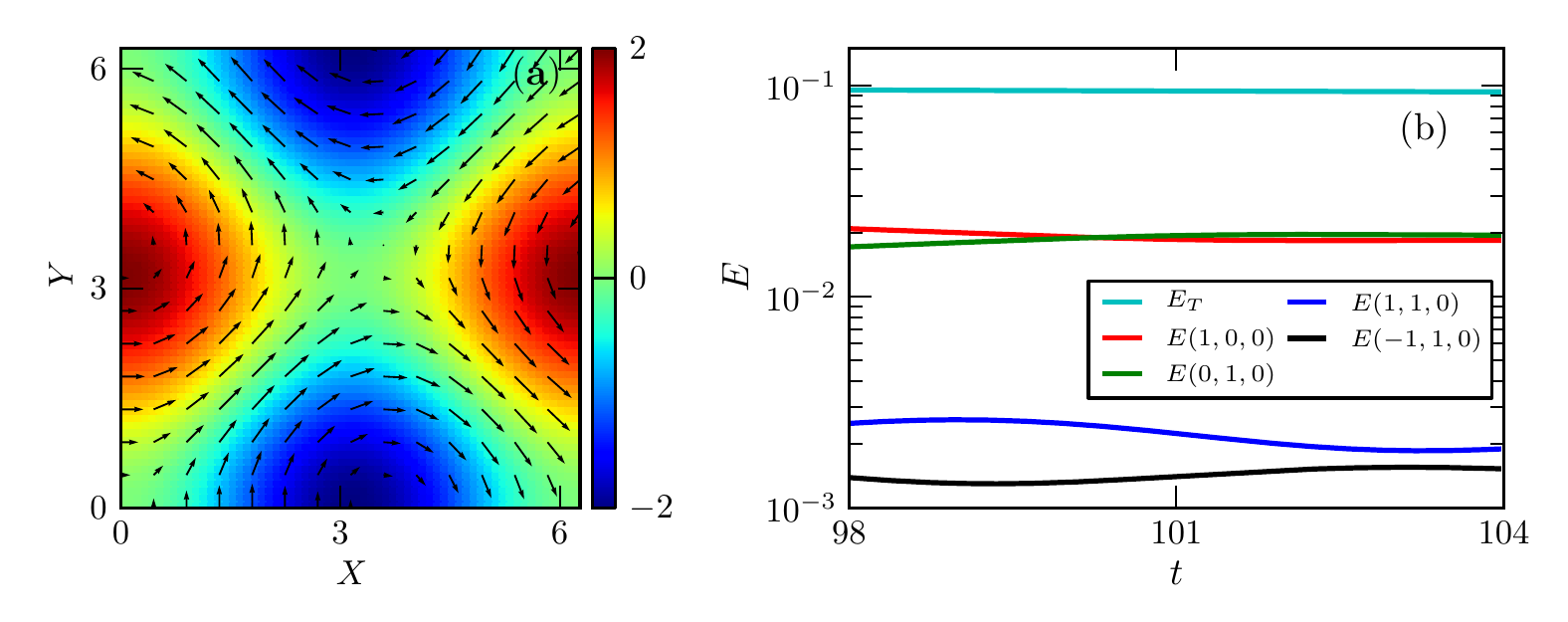}
\end{center}
\setlength{\abovecaptionskip}{0pt}
\caption{{ (a) For the 2D velocity field ${\bf u}(x,y) = \hat{x} \sin y + \hat{y} \sin x$ that corresponds to the Fourier modes $(1,0,0)$ and $(0,1,1)$, the vector plot of the velocity field superposed with the density plot of the vorticity field $\omega_z$.  (b) For the rotating simulation of Appendix~\ref{appA},  time series of the energies of the modes (1,0,0), (0,1,0), $(1,1,0)$, $(-1,1,0)$ and the total energy $E_T$. }}
\label{fig:8}
\end{figure}

\section{{Comparison between our model and that of Pao's model}}
\label{appC}

In this paper we argue that for strongly-rotating decaying turbulence, Equation~(\ref{eq:Ek_fit}) describes the energy spectrum in the intermediate and decaying range.  This is the prediction of our model based on the variable enstrophy flux.  However, it is important to compare it with the other models. In Figure~\ref{fig:9}, we plot $E(k)$ for the numerical data of $1024^3$  grid at $t=49$.  The plot also contains $E(k)$ predicted by Equation~(\ref{eq:Ek_fit})  and that by Pao's model (see Equation (\ref{eq:Pao_Ek})).  Clearly, our model performs better than Pao's model.  

We also remark that the predictions of Kraichnan's model given by Equation (\ref{eq:enstrophy_time_evolution}) is not suitable for our simulations because $\mathrm{Re} \gg 1$ for our flows.  Kraichnan's model assumes that the nonlinearity is absent, which is not the case for our system.  From  Equation (\ref{eq:enstrophy_time_evolution}) we deduce that the kinetic energy evolution in Kraichnan's model is
\begin{equation}
E(k,t) = E(k,0) \exp(-2\nu k^2 t).
\end{equation}
Thus, the evolution of $E(k)$ depends on the initial condition, and $E(k)$ quickly decays to zero.
}

\begin{figure}[htbp]
\begin{center}
\includegraphics[scale=1.0]{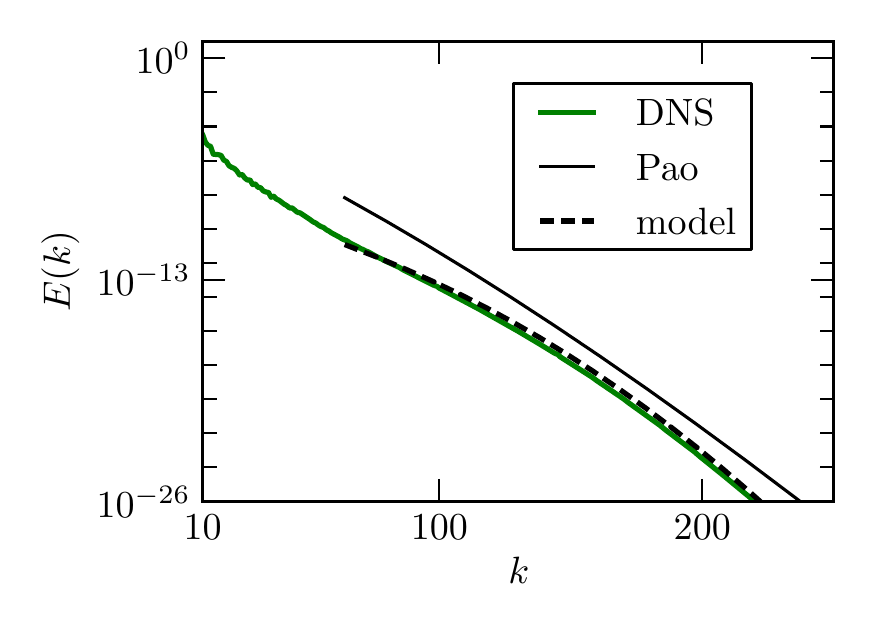}
\end{center}
\setlength{\abovecaptionskip}{0pt}
\caption{{ Plot of $E(k)$ for the 3D velocity field of $1024^3$ grid along with the predictions of our model and that of Pao.  This plot corresponds to the green curve of Figure~\ref{fig:6}.  }}
\label{fig:9}
\end{figure}

\section*{ Acknowledgments} 
We are thankful to Biplab Dutta for sharing his numerical results  that became the starting point of the present work. { We also thank the anonymous referees for useful suggestions.}  SC gratefully acknowledges financial support from the INSPIRE faculty fellowship (DST/INSPIRE/04/2013/000365) awarded by the INSA, India and DST, India. MKV thanks the Science and Engineering Research Board, India for the research grant (Grant No. SERB/F/3279), the Indian Space Research Organisation (ISRO), India for the research grant (Grant No. PLANEX/PHY/2015239), and the Department of Science and Technology, India (INT/RUS/RSF/P-03) and Russian Science Foundation Russia (RSF-16-41-02012) for the Indo-Russian project. The simulations were performed on the HPC2010 and HPC2013 and Chaos cluster of IIT Kanpur, India.


%

\end{document}